\documentclass[journal]{IEEEtran}
\usepackage{fancyhdr}
\usepackage{amsmath,amssymb}
\usepackage{amsfonts}
\usepackage[dvips]{graphicx}
\usepackage{dsfont}
\usepackage{comment}

\usepackage[hidelinks]{hyperref}    
\usepackage{graphicx}
\usepackage{color}
\usepackage{pgfplots}           
\pgfplotsset{compat=1.16}
\usepackage{enumerate}
\usepackage{graphics}
\usepackage{subfigure}
\usepackage{cite}
\usepackage{mathrsfs}
\usepackage{stfloats}

\usepgfplotslibrary{fillbetween}
\usetikzlibrary{patterns}
\usetikzlibrary {patterns,patterns.meta}

\usepackage{tikz}
\usepackage{mathdots}
\usepackage{yhmath}
\usepackage{cancel}
\usepackage{color}
\usepackage{siunitx}
\usepackage{array}
\usepackage{multirow}
\usepackage{textcomp}
\usepackage{gensymb}
\usepackage{tabularx}
\usepackage{booktabs}
\usepackage{graphicx}
\usepackage{xcolor}
\usepackage[utf8]{inputenc}
\usepackage[T1]{fontenc}

\newtheorem{remark}{Remark}

\renewcommand{\dbinom}[2]{\left(\!\!\begin{array}{c}{#1} \\ {#2} \end{array}\!\!\right)}
\let\oldbinom\binom
\renewcommand{\binom}[2]{\mathchoice{\dbinom{#1}{#2}}{\oldbinom{#1}{#2}}{\oldbinom{#1}{#2}}{\oldbinom{#1}{#2}}}
\newcommand{\bpsi}{\boldsymbol{\psi}}
\allowdisplaybreaks

\begin{document}
\title{Uplink RSMA for Pinching-Antenna Systems}
\author{
    Apostolos A. Tegos, Yue Xiao, Sotiris A. Tegos,~\IEEEmembership{Senior Member,~IEEE,}
    \\
    George K. Karagiannidis,~\IEEEmembership{Fellow,~IEEE,} and Panagiotis D. Diamantoulakis,~\IEEEmembership{Senior Member,~IEEE}
    \thanks{A. A. Tegos is with the Department of Electrical and Computer Engineering, Aristotle University of Thessaloniki, 54124 Thessaloniki, Greece (e-mail: apotegath@auth.gr).}
    \thanks{Yue Xiao, S. A. Tegos, G. K. Karagiannidis and A. D. Diamantoulakis are with the Department of Electrical and Computer Engineering, Aristotle University of Thessaloniki, 54124 Thessaloniki, Greece and with the Provincial Key Laboratory of Information Coding and Transmission, Southwest Jiaotong University, Chengdu 610031, China (e-mails: xiaoyue@swjtu.edu.cn, tegosoti@auth.gr, geokarag@auth.gr, padiaman@auth.gr).}
    \vspace{-4mm}
}

\maketitle
\begin{abstract}   
    One of the key goals of next-generation wireless networks is to adapt to changing conditions and meet the growing demand for reliable, high-capacity communications from emerging applications. Overcoming the limitations of conventional technologies, such as fixed antenna positions, is essential to achieving this objective because it mitigates the impact of path loss on the received signal and creates strong line-of-sight links, enhancing system performance. With this in mind, the newly proposed pinching antenna systems (PASs) are a promising solution for indoor applications because they can activate antennas across a waveguide deployed in a room, thus reducing the distance between the transmitter and receiver. In this paper, we investigate a two-user, two-pinching-antenna uplink PAS, in which the transmitters use rate splitting to create a more resilient framework than non-orthogonal multiple access (NOMA). For this network, we derive novel closed-form expressions for the outage probability. Numerical results validate these expressions, proving that the proposed rate-splitting multiple access (RSMA) scheme outperforms NOMA PAS.
\end{abstract}


\section{Introduction}
As next-generation wireless networks strive to meet the increasing demands of applications such as extended reality and advanced healthcare, it becomes clear that traditional fixed antennas are inadequate. These applications require ultra-reliable and high-capacity communication that can dynamically adjust to users with heterogeneous demands. To achieve these system characteristics, multiple antenna techniques have been extensively investigated in recent years. For example, massive multiple-input, multiple-output (MIMO) technology is attractive due to its ability to provide a higher degree of freedom. However, increasing the number of antennas in a fixed antenna system significantly reduces cost and energy efficiency. Considering this, programmable wireless environments (PWEs) have emerged as a fundamental solution in which wave propagation is dynamically adjusted to support different users' heterogeneous services. In particular, PWEs counter the stochastic nature of wireless communications by reconfiguring wireless propagation through software-defined processes to provide adaptive wireless networks.

\subsection{Literature Review}
In an effort to realize PWEs, researchers have made significant efforts to develop advanced reconfigurable technologies. One promising solution has been identified in the reconfigurable intelligent surface (RIS) technology, which manipulates incident electromagnetic waves using novel meta-materials \cite{RIS}. In more detail, by carefully placing and designing RISs, specific functionalities such as beam steering and diffusion can be exploited to meet the requirements of each user served by the network \cite{RISGeneral2}. Multiple advancements have been made on this topic, as researchers have investigated numerous RIS architectures with different capabilities, such as active RIS, which amplifies incoming signals \cite{RISActive}, light-emitting RIS, which simultaneously manipulates electromagnetic waves and utilizes optical signals for precise localization \cite{LERIS}, and zero-energy RIS, which harvests energy from incident signals for sustainable operation \cite{ZERIS}.   

Furthermore, other reconfigurable technologies such as fluid and movable antennas have been extensively investigated due to their ability to adapt both the receiver and the transmitter technologies depending on the conditions of the environment \cite{FA,MovAnt}. In particular, movable antennas are able to physically reposition themselves to improve channel conditions, while fluid antennas can alter their electromagnetic properties utilizing reconfigurable materials, such as liquid metals. Considering these unique capabilities, these technologies have been used in numerous scenarios such as security \cite{MovAntSec,FASec}, multiple access \cite{MovAntMA,FAMA}, and integrated security and communication (ISAC) \cite{MovAntISAC}. Therefore, these technologies can enhance channel conditions, support the PWE vision and thus be considered as promising solutions for next-generation networks.

The aforementioned techniques aim to support the PWE vision by improving the effective channel gain and mitigating the stochastic nature of wireless communication environments. Consequently, in line-of-sight (LoS) channels, the performance gains of these technologies over conventional fixed antenna systems diminish, proving a limitation in indoor scenarios. In such scenarios, path loss is a key aspect that limits system performance, especially in high-frequency bands, where its impact far exceeds that of signal degradation \cite{ChannelModelling}. Considering the fact that, in a RIS-assisted network, the transmitted signal first propagates from the transmitter to the RIS and then from the RIS to the receiver, it is evident that it suffers from double path loss \cite{PathLoss}. Furthermore, fluid and moving antennas cannot reduce path loss, since their positional adjustments are limited to a few wavelengths, resulting in minor improvements.

These limitations of existing solutions for PWE motivate the exploration of pinching-antenna systems (PASs), an innovative concept introduced by DOCOMO in 2022 that uses dielectric waveguides to guide electromagnetic waves at high frequencies \cite{DOCOMO,Pinching}. In PASs, the waveguide, connected to the access point (AP), is positioned along the edge of the ceiling. Pinching antennas (PAs), which are the radiating elements, are activated by applying dielectric particles at any desired point across the waveguide \cite{ding2025flexible}. This unique attribute allows for flexible antenna placement along the waveguide, thus minimizing the distance between the PA and the user. Furthermore, it can facilitate establishing LoS links in obstructed environments by adequately placing the PAs without requiring additional hardware, making them a practical solution for reconfigurable networks. 

Considering their multiple benefits, researchers have recently investigated PASs to showcase their capabilities and compare their performance to that of existing technologies. Specifically, in \cite{ding2025flexible}, a PAS for indoor wireless communication was introduced, investigating both orthogonal multiple access (OMA) and non-orthogonal multiple access (NOMA) for downlink transmissions. Closed-form expressions were derived for the case of a single PA and a single waveguide, which were then extended for multiple PAs and multiple waveguides. Furthermore, in \cite{Tyrovolas}, the outage probability (OP) and the average rate for a single PA serving a single user were investigated, taking into account waveguide attenuation.  The maximization of the minimum data rate for the uplink and the data rate for the downlink scenario was studied in \cite{RateUP,RateDown}, respectively. In \cite{RateUP}, a method to successively optimize the position of the PAs and the resource allocation was proposed, while in \cite{RateDown}, the optimization problem considered moving the PAs to minimize the path loss while maintaining constructive interference to the served user. In \cite{Thrassos}, an orthogonal frequency division multiple access (OFDMA)-based framework is introduced to mitigate inter-symbol interference in PAS, which is used to maximize user fairness. Finally, PAS for physical layer security is investigated in \cite{sunsec,Boz}. Specifically, in \cite{sunsec}, the optimal pinching beamforming is studied for the scenarios with single or multiple legitimate users and eavesdroppers. In \cite{Boz}, the maximization of secrecy rate utilizing appropriate beamforming and artificial noise is examined in single and multiple waveguide scenarios. 

\subsection{Motivation and Contribution}
Existing works have demonstrated the potential of PASs and the need for further research. However, most of these studies have focused on the downlink scenario, while the uplink has received limited attention. Specifically, the uplink has been examined in \cite{RateUP} and \cite{Uplink}, where the optimal positions of the PAs, and closed-form expressions for the ergodic rate were derived. However, multiple approximations were used to derive these expressions. Additionally, the scenario of multiple PAs serving multiple users has not been investigated, despite being the most common scenario in practice. 

In addition, existing works have considered either NOMA or OMA when investigating multiple access. However, rate-splitting multiple access (RSMA) \cite{RSMA} has recently received extensive attention in both academia and industry, since it provides a more resilient transmission framework compared to NOMA. In uplink RSMA, users split their messages into streams and transmit each stream at a specific power level. NOMA can be achieved as a special case of RSMA by choosing the power levels adequately. At the receiver, the AP employs successive interference cancellation (SIC) to decode the transmitted messages. Considering the above, the contribution of this work can be summarized as follows:
\begin{itemize}
    \item We consider an uplink PAS consisting of two PAs, which receive messages from two users. Each user is located in a different room, thus its messages are received by only one PA. We assume that the users employ RSMA to transmit their data, while the decoding takes place at the AP at the feed of the waveguide. 
    \item For such a network, we derive closed-form expressions for the OP of each message. These novel expressions are derived without approximations and provide useful practical and theoretical insights into the optimal system design. 
    \item Simulation results validate the theoretical analysis, demonstrating that the proposed system outperforms conventional fixed antenna systems by reducing the path loss affecting the transmitted signals. Furthermore, it is shown that, by optimizing the additional degrees of freedom provided by RSMA, it can outperform NOMA. Most importantly, it is not limited by OP floors. 
\end{itemize}

\section{System Model}
\begin{figure}
\centering


\begin{tikzpicture}[x=0.75pt,y=0.75pt,yscale=-0.65,xscale=0.53]

\draw   (216.29,46.11) -- (653,46.11) -- (465.84,302) -- (29.13,302) -- cycle ;
\draw  [dash pattern={on 4.5pt off 4.5pt}]  (29.13,302) -- (232.22,25.42) ;
\draw [shift={(234,23)}, rotate = 126.29] [fill={rgb, 255:red, 0; green, 0; blue, 0 }  ][line width=0.08]  [draw opacity=0] (8.93,-4.29) -- (0,0) -- (8.93,4.29) -- cycle    ;
\draw  [dash pattern={on 4.5pt off 4.5pt}]  (131.57,162.5) -- (425.97,161.93) -- (597.33,161.34) ;
\draw [shift={(600.33,161.33)}, rotate = 179.8] [fill={rgb, 255:red, 0; green, 0; blue, 0 }  ][line width=0.08]  [draw opacity=0] (8.93,-4.29) -- (0,0) -- (8.93,4.29) -- cycle    ;
\draw  [fill={rgb, 255:red, 235; green, 232; blue, 233 }  ,fill opacity=1 ] (568.26,52.43) -- (131.26,53.57) -- (131.28,59.81) -- (568.28,58.67) -- cycle ;
\draw  [dash pattern={on 4.5pt off 4.5pt}]  (131.28,59.67) -- (131.57,162.5) ;
\draw  [fill={rgb, 255:red, 128; green, 128; blue, 128 }  ,fill opacity=1 ] (209.76,65.06) -- (200.96,50.56) -- (218.28,50.41) -- cycle ;
\draw  [dash pattern={on 4.5pt off 4.5pt}]  (568.28,58.67) -- (568.57,161.5) ;
\draw  [dash pattern={on 4.5pt off 4.5pt}]  (435.84,47.02) -- (375.04,126.37) -- (256.34,281.3) -- (239.2,303.91)(433.46,45.19) -- (372.66,124.54) -- (253.96,279.48) -- (236.8,302.09) ;
\draw  [dash pattern={on 4.5pt off 4.5pt}]  (345.28,58.67) -- (345.57,161.5) ;
\draw  [fill={rgb, 255:red, 128; green, 128; blue, 128 }  ,fill opacity=1 ] (468.76,65.06) -- (459.96,50.56) -- (477.28,50.41) -- cycle ;
\draw (181.75,187) node  {\includegraphics[width=28.13pt,height=22.5pt]{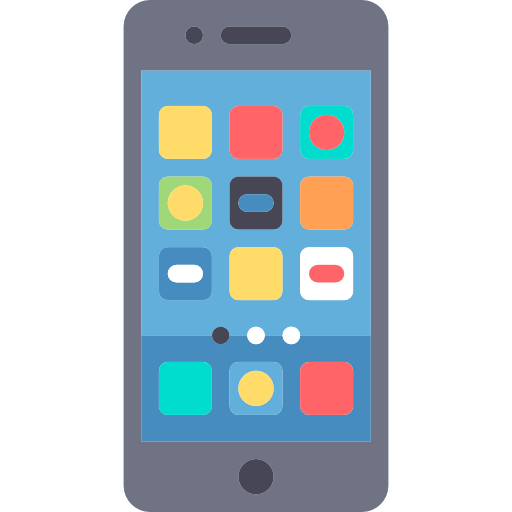}};
\draw (453.75,205) node  {\includegraphics[width=28.13pt,height=22.5pt]{phone_icon.png}};
\draw (345.34,39.46) node  {\includegraphics[width=20.16pt,height=18.69pt]{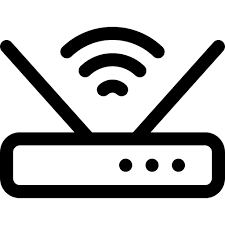}};

\draw (349,169.99) node [anchor=north west][inner sep=0.75pt]    {$( 0,0,0)$};
\draw (609.2,137.17) node [anchor=north west][inner sep=0.75pt]  [font=\large]  {$x$};
\draw (209,7.72) node [anchor=north west][inner sep=0.75pt]    {$y$};
\draw (115,106.37) node [anchor=north west][inner sep=0.75pt]    {$d$};
\draw (68.4,34.77) node [anchor=north west][inner sep=0.75pt]   [align=left] {waveguide};
\draw (115.7,206.89) node [anchor=north west][inner sep=0.75pt]    {$\boldsymbol{\psi }_{\text{U} ,1} =\left( x_{U,1} ,y_{U,1} ,0\right)$};
\draw (293,11) node [anchor=north west][inner sep=0.75pt]   [align=left] {AP};
\draw (52,277) node [anchor=north west][inner sep=0.75pt]   [align=left] {{\fontfamily{ptm}\selectfont Room 1}};
\draw (380,277) node [anchor=north west][inner sep=0.75pt]   [align=left] {{\fontfamily{ptm}\selectfont Room 2}};
\draw (357.7,228.89) node [anchor=north west][inner sep=0.75pt]    {$\boldsymbol{\psi }_{U,2} =\left( x_{U,2} ,y_{U,2} ,0\right)$};
\draw (155,69.4) node [anchor=north west][inner sep=0.75pt]    {$\boldsymbol{\psi }_{P,1} =( x_{P,1} ,0,d)$};
\draw (427,22.4) node [anchor=north west][inner sep=0.75pt]    {$\boldsymbol{\psi }_{P,2} =( x_{P,1} ,0,d)$};

\end{tikzpicture}
\vspace{-4mm}
\caption{System model.}
\vspace{-4mm}
\label{System model}
\end{figure}
We assume a wireless uplink PAS consisting of an AP at $(0,0,d)$, two PAs, denoted by $P_1$ and $P_2$ and two single-antenna devices, denoted by $U_1$ and $U_2$. It is assumed that $U_1$ and $U_2$ are uniformly distributed in room 1 and room 2, respectively, where both rooms are rectangular lying in the $x-y$ plane with sides $D_x$ and $D_y$. Without loss of generality, the waveguide is considered to be parallel to the $x$-axis, with its height denoted by $d$ and its length equal to $2D_x$ in order to cover both rooms, as seen in Fig. \ref{System model}. Considering that the positions of the PAs and users are given respectively as $\bpsi_{P,i} = (x_{P,i},0,d)$, $\bpsi_{U,i} = (x_{U,i},y_{U,i},0)$, $i \in \{1,2\}$ and that $P_1$ receives data from $U_1$ and $P_2$ receives data from $U_2$, while a wall between the rooms does not allow transmissions to penetrate to the other room, we set $x_{P,i}=x_{U,i}$ to optimize the performance of the system. Therefore, the channel between PA $i$ and the respective user can be expressed as
\begin{equation} \small
    h_{1,i} = \frac{\sqrt{\eta}e^{-j\frac{2\pi}{\lambda}\|\bpsi_{P,i}-\bpsi_{U,i}\|}}{\|\bpsi_{P,i}-\bpsi_{U,i}\|},
\end{equation}
with $\eta = \frac{\lambda^2}{16\pi^2}$ representing the path loss at a reference distance of 1m, $\lambda$ denoting the free-space wavelength of the signal, $j$ is the imaginary unit, and $\|\cdot\|$ expressing the Euclidean norm. Furthermore, since the decoding takes place in the AP, which is considered to be at the wall between the rooms, a phase shift is induced to the received signal due to the propagation distance in the waveguide. Specifically, the interaction of the signal with the core and material within the waveguide reduces its phase velocity, characterized by the effective refractive index $n_{e}$, which determines the guided wavelength as $\lambda_g=\frac{\lambda}{n_{e}}$.  Consequently, the received signal is affected by a phase shift, given by
\begin{equation} \small
    h_{2,i} = e^{-j\frac{2\pi}{\lambda_g}\|\bpsi_{P,i}-\bpsi_0\|},
\end{equation}
with $\bpsi_0=(0,0,d)$. In this work, we consider that RSMA is employed by the users and without loss of generality, we assume that $U_1$ is the one performing message splitting, and thus the message received at the BS is given as
\begin{equation} \small
    \begin{aligned}
        y =& \sqrt{\alpha P_1}h_{1,1}h_{2,1}x_{1a} + \sqrt{(1-\alpha)P_1}h_{1,1}h_{2,1}x_{2a} \\
        &+ \sqrt{P_2}h_{1,2}h_{2,2}x_{b} + n_b,
    \end{aligned}
\end{equation}
where $\alpha$ and $1-\alpha$ are the power allocation coefficients for messages $x_{1a}$ and $x_{2a}$, respectively. Furthermore, $P_1$ and $P_2$ are the transmit powers of each user, $n_b$ is the additive white Gaussian noise (AWGN) with zero mean and variance $\sigma^2$ at the BS. It has been proven that for uplink RSMA to achieve the capacity region, when $K$ users are transmitting, only $K-1$ users need to split their messages, while the decoding order of the transmitted messages that achieves this is ($x_{1a}$, $x_{b}$, $x_{2a}$). Thus, considering this decoding order, the received signal-to-interference-plus-noise ratio (SINR) at the BS for detecting message $x_{1a}$ is given by \eqref{SINRstart} at the top of this page. 
\begin{figure*}
    \begin{equation} \small\label{SINRstart}
        \gamma_{1a} = \frac{\frac{\alpha\eta P_1\bigg | e^{-j\left(\ \frac{2\pi}{\lambda}\|\bpsi_{P,1}-\bpsi_{U,1}\|+\frac{2\pi}{\lambda_g}\|\bpsi_{P,1}-\bpsi_0\right)}\bigg|^2}{\|\bpsi_{P,1}-\bpsi_{U,1}\|^2}}{\frac{(1-\alpha)\eta P_1\bigg | e^{-j\left(\ \frac{2\pi}{\lambda}\|\bpsi_{P,1}-\bpsi_{U,1}\|+\frac{2\pi}{\lambda_g}\|\bpsi_{P,1}-\bpsi_0\right)}\bigg|^2}{\|\bpsi_{P,1}-\bpsi_{U,1}\|^2}+\frac{\eta P_2\bigg | e^{-j\left(\ \frac{2\pi}{\lambda}\|\bpsi_{P,2}-\bpsi_{U,2}\|+\frac{2\pi}{\lambda_g}\|\bpsi_{P,2}-\bpsi_0\right)}\bigg|^2}{\|\bpsi_{P,2}-\bpsi_{U,2}\|^2}+\sigma^2}
    \end{equation}
    \hrulefill
\end{figure*}
Considering $|e^{-jx}|=1$, $\gamma_1=\frac{P_1}{\sigma^2}$, $\gamma_2=\frac{P_2}{\sigma^2}$ and  
    $\|\bpsi_{P,i}-\bpsi_{U,i}\| = \sqrt{y_{U,i}^2+d^2}$,
\eqref{SINRstart} can be rewritten as
\begin{equation} \small \label{SNRfinal1a}
    \gamma_{1a} = \frac{\alpha\eta\gamma_1\left(y_{U,2}^2+d^2\right)}{(1-\alpha)\gamma_1\eta(y_{U,2}^2+d^2)+(\eta\gamma_2 + y_{U,2}^2+d^2)(y_{U,1}^2+d^2)} .
\end{equation}
Similarly, the SINRs for messages $x_{b}$ and $x_{2a}$ are given by
\begin{equation} \small \label{SNRxb}
    \gamma_b = \frac{\eta\gamma_2(y_{U,1}^2+d^2)}{((1-\alpha)\eta\gamma_1+y_{U,1}^2+d^2)(y_{U,2}^2+d^2)},
\end{equation}
and
\begin{equation} \small \label{SNRx2a}
    \gamma_{2a} = \frac{(1-\alpha)\eta\gamma_1}{y_{U,1}^2+d^2}.
\end{equation}
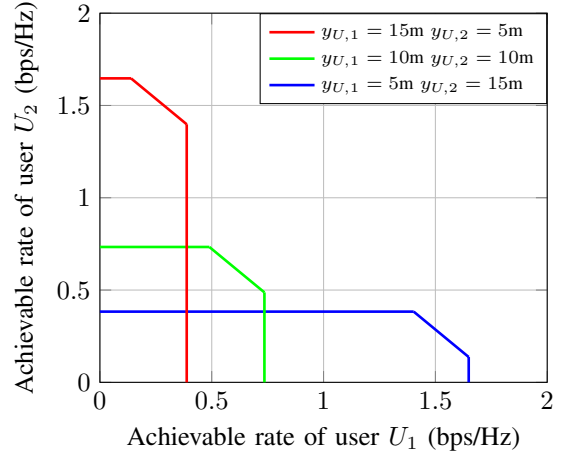
\begin{figure}
        \centering
        \begin{tikzpicture}
            \begin{axis}[
                width = 0.85\linewidth,
                xlabel = {Achievable rate of user $U_1$ (bps/Hz)},
                ylabel = {Achievable rate of user $U_2$ (bps/Hz)},
                ymin = 0,
                ymax = 2,
                xmin = 0,
                xmax = 2,
                grid = major,
                legend entries = {{$y_{U,1}=15$m $y_{U,2}=5$m},{$y_{U,1}=10$m $y_{U,2}=10$m},{$y_{U,1}=5$m $y_{U,2}=15$m},{},{},{},{},{},{},{}},
                legend cell align = {left},
                legend style = {font = \scriptsize},
                legend style={at={(1,1)},anchor=north east}
                ]
                \addplot[
                red,
                mark = square,
                mark repeat = 2,
                mark size = 3,
                mark phase = 0,
                no marks,
                line width = 1pt
                ]
                table {Data/CapacityRegion/CCR115R25.dat};   
                \addplot[
                green,
                mark = triangle,
                mark repeat = 2,
                mark size = 3,
                mark phase = 0,
                no marks,
                line width = 1pt
                ]
                table {Data/CapacityRegion/CCR110R210.dat};
                \addplot[
                blue,
                mark = o,
                mark repeat = 2,
                mark size = 3,
                mark phase = 0,
                no marks,
                line width = 1pt
                ]
                table {Data/CapacityRegion/CCR15R215.dat};
                \addplot[
                blue,
                mark = square,
                mark repeat = 2,
                mark size = 3,
                mark phase = 0,
                no marks,
                line width = 1pt
                ]
                table {Data/CapacityRegion/x1.dat};   
                \addplot[
                blue,
                mark = triangle,
                mark repeat = 2,
                mark size = 3,
                mark phase = 0,
                no marks,
                line width = 1pt
                ]
                table {Data/CapacityRegion/y1.dat};
                \addplot[
                green,
                mark = o,
                mark repeat = 2,
                mark size = 3,
                mark phase = 0,
                no marks,
                line width = 1pt
                ]
                table {Data/CapacityRegion/x2.dat};
                \addplot[
                green,
                mark = square,
                mark repeat = 10,
                mark size = 3,
                mark phase = 0,
                no marks,
                line width = 1pt
                ]
                table {Data/CapacityRegion/y2.dat};   
                \addplot[
                red,
                mark = triangle,
                mark repeat = 10,
                mark size = 3,
                mark phase = 0,
                no marks,
                line width = 1pt
                ]
                table {Data/CapacityRegion/x3.dat};
                \addplot[
                red,
                mark = o,
                mark repeat = 10,
                mark size = 3,
                mark phase = 0,
                no marks,
                line width = 1pt
                ]
                table {Data/CapacityRegion/y3.dat};
            \end{axis}
        \end{tikzpicture}
        \vspace{-4mm}
        \caption{Capacity region of PAS RSMA.}
        \vspace{-2mm}
        \label{CapReg}
    \end{figure} 
    
Fig. \ref{CapReg} shows the capacity region of the proposed protocol for different user locations. It should be noted that the location of the users along the $x$-axis, i.e., the axis on which the waveguide is deployed, does not affect the performance of the system because of the PAs' unique ability to adjust their position on the waveguide to minimize the path loss. As previously mentioned, RSMA is a capacity-achieving protocols, while by varying the value of $\alpha \in [0,1]$ every point on the diagonal line can be achieved. Note that NOMA can achieve only the corner points of this line. Regarding the connection of this figure to the following outage analysis, if the rate threshold is lower than each user's achievable rate, the transmitted messages will be correctly decoded. However, if the rate threshold is higher than the achievable rates, i.e., if it is outside the capacity region, an outage will occur.
\section{Outage Probability Analysis}
In this section, the derivation of closed-form expressions for the OP of all messages is presented. OP is an important metric for evaluating system performance, when the data rate is fixed. Being able to reliably decode the transmitted messages is crucial in multiple real world scenarios. To derive the OP, we consider predefined thresholds, given as $\theta_{11}=2^{\beta R_1}-1$, $\theta_{12}=2^{(1-\beta)R_1}-1$, and $\theta_2=2^{R_2}-1$ for messages ($x_{1a}$, $x_{2a}$, $x_{b}$), where $R_1$ and $R_2$ are the target rates for each user, and $\beta \in [0,1]$ denotes the target rate factor. To keep the closed-form expressions as compact as possible, we define the necessary functions in Table \ref{HelpingFunctions} and \eqref{Phi1}-\eqref{Phi4} at the top of the next page.

\begin{figure*}
    \begin{equation} \small \label{Phi1}
        \begin{aligned}
            \Phi_1(x,y) =  &\frac{4}{D_y^2}\sqrt{\frac{Q_2}{S_2}}\left(\frac{\frac{P_2}{Q_2}}{\sqrt{\frac{R_2}{S_2}}}F\left(f_1(x),T_2\right)-\sqrt{\frac{R_2}{S_2}}E\left(f_1(x),T_2\right)+f_2(x)  -\frac{\frac{P_2}{Q_2}}{\sqrt{\frac{R_2}{S_2}}}F\left(f_1(y),T_2\right)-\sqrt{\frac{R_2}{S_2}}E\left(f_1(y),T_2\right)+f_2(y)\right)\\
            &+\frac{4}{D_y^2}\sqrt{\frac{Q_1}{S_1}}\left(\sqrt{\frac{P_1}{S_1}+\frac{R_1}{S_1}}E\left(f_3(y),T_1\right)-\sqrt{\frac{P_1}{S_1}+\frac{R_1}{S_1}}E\left(f_3(x),T_1\right)\right)
        \end{aligned}
    \end{equation}
    \hrulefill
\end{figure*}
\begin{figure*}
    \begin{equation} \small \label{Phi2}
        \Phi_2(x,y) \!=\! 
        \frac{4}{D_y^2}\sqrt{\frac{Q_2}{S_2}}\left(\frac{\frac{P_2}{Q_2}}{\sqrt{\frac{R_2}{S_2}}}F\left(f_1(x),T_2\right)-\sqrt{\frac{R_2}{S_2}}E\left(f_1(x),T_2\right)+f_2(x) -\frac{\frac{P_2}{Q_2}}{\sqrt{\frac{R_2}{S_2}}}F\left(f_1(y),T_2\right)-\sqrt{\frac{R_2}{S_2}}E\left(f_1(y),T_2\right)+f_2(y)\right)   
    \end{equation}
    \hrulefill
\end{figure*}
\begin{figure*}
    \begin{equation} \small \label{Phi3}
        \Phi_3(x,y,z) = \frac{4z}{D_y^2}(x-y)-\frac{4}{D_y^2}\sqrt{\frac{Q_1}{S_1}}\sqrt{\frac{P_1}{Q_1}+\frac{R_1}{S_1}}\left(E\left(f_3(y),T_1\right)-E\left(f_3(x),T_1\right)\right)
    \end{equation}
    \hrulefill
\end{figure*}
\begin{figure*}
    \begin{equation} \small \label{Phi4}
        \Phi_4(x,y) = \frac{4}{D_y^2}\sqrt{\frac{Q_1}{S_1}}\sqrt{\frac{P_1}{Q_1}+\frac{R_1}{S_1}}\left(E\left(f_3(y),T_1\right)-E\left(f_3(x),T_1\right)\right)
    \end{equation}
    \hrulefill
\end{figure*}

\begin{table}
    \centering
    \caption{Used Functions}
    \begin{tabular}{c|c}
        \hline \hline
        Term    & Expression  \\
        \hline \hline
        $C_3$ &
        $\frac{\eta\gamma_1\left(\alpha-(1-\alpha)\theta_{11}\right)-\theta_{11}d^2}{\theta_{11}}$ 
        \\ \hline
        $C_4$ &
        $\frac{\left(d^2+\frac{D_y^2}{4}\right)\eta\gamma_1\left(\alpha-(1-\alpha)\theta_{11}\right)}{\theta_{11}\left(\eta\gamma_2+d^2+\frac{D_y^2}{4}\right)}-d^2$
        \\ \hline
        $C_5$ &
        $d^2\left(\frac{\eta\gamma_1\alpha(1+\theta_{11})-\left(d^2+\eta(\gamma_1+\gamma_2)\right)\theta_{11}}{\left(d^2+\eta\gamma_2\right)\theta_{11}}\right)$ 
        \\ \hline
        $C_6$ &
        $d^2\left(\frac{\eta\gamma_2-d^2\theta_2-(1-\alpha)\eta\gamma_1\theta_2}{d^2\theta_2-\eta\gamma_2}\right)$
        \\ \hline
        $C_7$ &
        $\frac{d^2\left(4\eta\gamma_2-D_y^2\theta_2-4(1-\alpha)\eta\gamma_1\theta_2\right)-4d^4\theta_2-(1-\alpha)D_y^2\eta\gamma_1\theta_2}{\left(4d^2+D_y^2\right)\theta_2-4\eta\gamma_2}$
        \\ \hline
        $C_8$ &
        $\frac{(1-\alpha)\eta\gamma_1}{\theta_{12}}-d^2$
        \\ \hline
        $C_9$ &
        $\left(\alpha+\frac{\alpha}{\theta_{11}(1+\theta_2)}-1\right)\eta\gamma_1-d^2$
        \\ \hline
        $Q_1$ &
        $\eta\gamma_2\theta_{11}+\theta_{11}d^2$
        \\ \hline
        $P_1$ &
        $\theta_{11}d^4+d^2\theta_{11}\eta\gamma_2-d^2\eta\gamma_1(\alpha-(1-\alpha)\theta_{11})$
        \\ \hline
        $R_1$ &
        $\eta\gamma_1(\alpha-(1-\alpha)\theta_{11})-d^2\theta_{11}$
        \\ \hline
        $S_1$ &
        $\theta_{11}$
        \\ \hline
        $T_1$ &
        $\left(\sqrt{\frac{R_1}{S_1}}/\sqrt{\frac{P_1}{Q_1}+\frac{R_1}{S_1}}\right)^2$
        \\ \hline
        $Q_2$ &
        $\eta\gamma_2-d^2\theta_2$ 
        \\ \hline
        $P_2$ &
        $d^2\eta\gamma_2-d^2\eta\gamma_1\theta_2(1-\alpha)-d^4\theta_2$
        \\ \hline
        $R_2$ &
        $\eta\gamma_1\theta_2(1-\alpha)+b^2\theta_2$
        \\ \hline
        $S_2$ &
        $\theta_2$ 
        \\ \hline
        $T_2$ &
        $\left(\sqrt{\frac{R_2}{S_2}}/\sqrt{\frac{P_2}{Q_2}+\frac{R_2}{S_2}}\right)^2$
        \\ \hline
        $f_1(x)$ &
        $\arctan \left(x/\sqrt{\frac{P_2}{Q_2}}\right)$  
        \\ \hline
        $f_2(x)$ &
        $x\sqrt{\frac{\frac{R_2}{S_2}+x^2}{\frac{P_2}{Q_2}+x^2}}$
        \\ \hline
        $f_3(x)$ &
        $\arccos \left(x/\sqrt{\frac{R_1}{S_1}}\right)$
        \\ \hline \hline
    \end{tabular}
    
    \label{HelpingFunctions}
    \vspace{-3mm}
\end{table}

\subsection{Outage probability for message $x_{1a}$}
In this subsection, closed-form expressions are provided for the OP of message $x_{1a}$.
\begin{table*}
    \caption{Expressions for OP $x_{1a}$}
    \label{OP1a}
    \centering
    \begin{tabular}{c|c|c|c}
        \hline \hline
        \multicolumn{3}{c|}{Conditions}
        & Expression
        \\ \hline \hline
        $C_3 < 0$ &
        &
        &
        1
        \\ \hline
        \multirow{3}{*}[0cm]
        {$C_3 \geq 0 ,C_3 < \frac{D_y^2}{4}$} &
        \multirow{2}{*}[0cm]
        {$C_4 \geq 0 ,C_4 < \frac{D_y^2}{4}$} &
        $C_5 \geq 0 ,C_5 < \frac{D_y^2}{4}$ &
        $\Phi_4\left(\sqrt{C_4},\sqrt{C_5}\right)+\frac{2}{D_y}\left(\frac{D_y}{2}-\sqrt{C_4}\right)$
        \\ \cline{3-4}
        &
        &
        $C_5 < 0$ &
        $\Phi_4\left(\sqrt{C_4},0\right)+\frac{2}{D_y}\left(\frac{D_y}{2}-\sqrt{C_4}\right)$
        \\ \cline{2-4}
        &
        $C_4 < 0$ &
        &
        1
        \\ \hline
        \multirow{6}{*}[0cm]{$C_3\geq\frac{D_y^2}{4}$} &
        \multirow{3}{*}[0cm]{$C_4\geq\frac{D_y^2}{4}$} &
        $C_5\geq\frac{D_y^2}{4}$ &
        0
        \\ \cline{3-4}
        &
        &
        $C_5\geq0 ,C_5<\frac{D_y^2}{4}$ &
        $\Phi_4\left(\frac{D_y}{2},\sqrt{C_5}\right)$ 
        \\ \cline{3-4}
        &
        &
        $C_5<0$ &
        $\Phi_4\left(\frac{D_y}{2},0\right)$
        \\ \cline{2-4}
        &
        \multirow{2}{*}[0cm]{$C_4\geq0 , C_4<\frac{D_y^2}{4}$} &
        $C_5\geq0 , C_5<\frac{D_y^2}{4}$ &
        $\Phi_4\left(\sqrt{C_4},\sqrt{C_5}\right)+\frac{2}{D_y}\left(\frac{D_y}{2}-\sqrt{C_4}\right)$ 
        \\ \cline{3-4}
        &
        &
        $C_5<0$ &
        $\Phi_4\left(\sqrt{C_4},0\right)+\frac{2}{D_y}\left(\frac{D_y}{2}-\sqrt{C_4}\right)$ 
        \\ \cline{2-4}
        &
        $C_4<0$ &
        &
        1
        \\ \hline \hline
    \end{tabular}
\end{table*}
\begin{IEEEproof}
    In the proposed scheme, an outage for message $x_{1a}$ occurs when the received SINR for this message, i.e., $\gamma_{1a}$, as given by \eqref{SNRfinal1a}, is lower than the corresponding threshold $\theta_{11}$. Starting from 
    \begin{equation} \small
        P_{o,x_{1a}} = \Pr\left(\gamma_{1a} < \theta_{11}\right)
    \end{equation} and, after some algebraic manipulation, we get
    \begin{equation} \small \label{Pout1A}
        \begin{aligned} 
            P_{o,x_{1a}} = & \Pr\left(\left(y_{U,2}^2+d^2\right)\left(\eta\gamma_1(\alpha-(1-\alpha)\theta_{11}) \right. \right.\\&\left.\left.- \theta_{11}\left(y_{U,1}^2+d^2\right)\right) \leq \eta\gamma_2\left(y_{U,1}^2+d^2\right)\theta_{11}\right).
        \end{aligned}
    \end{equation}
    Taking into account that the right-hand side of the inequality is positive, the conditions with regard to $C_3$ can be derived. When $C_3 \geq 0$, it occurs that an outage occurs when $y_{U,2}^2\leq C_1$, where $C_1 = \frac{\eta\gamma_2\left(y_{U,1}^2+d^2\right)\theta_{11}}{\eta\gamma_1(\alpha-(1-\alpha)\theta_{11})-\theta_{11}\left(y_{U,1}^2+d^2\right)}-d^2$. Furthermore, by considering that $C_1$ should be positive, the condition that $y_{U,1}^2\geq C_5$ is derived. It should be highlighted that the random variables $y_{U,1}$ and $y_{U,2}$ are independent and uniformly distributed in $[-\frac{D_y}{2},\frac{D_y}{2}]$, thus the joint probability density function $f_{y_{U,1},y_{U,2}}(y_{U,1},y_{U,2})=\frac{1}{D_y^2}$, which necessitates an additional comparison between $C_1$ and $\frac{D_y}{2}$ from which it is derived when $y_{U,1}\geq \sqrt{C_4}$, $\sqrt{C_1} \geq \frac{D_y}{2}$. Indicatively, the derivation of the final expressions for one of the cases are presented below. When $C_3\geq \frac{D_y^2}{4}$, $C_4\geq0$, $C_4\leq \frac{D_y^2}{4}$, $C_5\geq0$, and $C_5\leq \frac{D_y^2}{4}$ and considering the symmetry of the system model, the OP is given by
    \begin{equation} \small \small
        \begin{aligned}
            P_{o,x_{1a}} \! = \! 4\!\underbrace{\int_{\!\sqrt{C_5}}^{\!\sqrt{C_4}} \!\! \int_0^{\sqrt{C_1}} \!\!\! \frac{1}{D_y^2}\mathrm{d} y_{U,2}\mathrm{d} y_{U,1}}_{I_1} 
            \!+ 4 \!\underbrace{\int_{\!\sqrt{C_4}}^{\frac{D_y}{2}} \! \int_{0}^{\frac{D_y}{2}} \!\!\! \frac{1}{D_y^2}\mathrm{d} y_{U,2}\mathrm{d} y_{U,1}}_{I_2} \! .
        \end{aligned}
    \end{equation}
    The result of the second integral, i.e., $I_2$, is straightforward, the first integral, however, assumes the form of 
    \begin{equation} \small
        I_1 \!=\!\frac{4}{D_y^2} \!\int_{\!\sqrt{C_5}}^{\!\sqrt{C_4}}\!\sqrt{\!\frac{\eta\gamma_2\left(y_{U,1}^2+d^2\right)\theta_{11}}{\eta\gamma_1(\alpha\!-\!(1\!-\!\alpha)\theta_{11})\!-\!\theta_{11}\left(y_{U,1}^2+d^2\right)}\!-\!d^2}\mathrm{d}y_{U,1},
    \end{equation}
    which after some algebraic manipulations results in 
    \begin{equation} \small \label{I1start}
        I_1 = \frac{4}{D_y^2}\int_{\sqrt{C_5}}^{\sqrt{C_4}}\sqrt{\frac{Q_1y_{U,1}^2+P_1}{R_1-S_1y_{U,1}^2}}\mathrm{d}y_{U,1},
    \end{equation}
    where $Q_1$, $P_1$, $S_1$, and $T_1$ are provided in Table \ref{HelpingFunctions}. To derive a closed-form expression for \eqref{I1start}, \cite[(3.169.4)]{gradshteyn2014table} is utilized. However, in this formula the upper limit of the integral should be equal to the term in the denominator, thus \eqref{I1start} is transformed into 
    \begin{equation} \small\label{I1final}
        \begin{aligned}
            I_1 = &\frac{4}{D_y^2}\sqrt{\frac{Q_1}{S_1}}\left(\int_{\sqrt{C_5}}^{\frac{R1}{S1}}\!\sqrt{\frac{y_{U,1}^2+\frac{P_1}{Q_1}}{\frac{R_1}{S_1}-y_{U,1}^2}}\mathrm{d}y_{U,1} \right.\\&\left.-\int_{\sqrt{C_4}}^{\frac{R1}{S1}}\sqrt{\frac{y_{U,1}^2+\frac{P_1}{Q_1}}{\frac{R_1}{S_1}-y_{U,1}^2}}\mathrm{d}y_{U,1}\right).
        \end{aligned}
    \end{equation}
    Although the limits of the integral have been altered, the various cases have been formulated accordingly so that the sign of the radicand remains positive and that $\frac{R_1}{S_1} \geq \sqrt{C_5}$ and $\frac{R_1}{S_1} \geq \sqrt{C_4}$. Invoking \cite[(3.169.4)]{gradshteyn2014table} in \eqref{I1final}, the final expressions can be derived. Following a similar procedure, the rest of the cases in Table \ref{OP1a} can be calculated.
\end{IEEEproof}
\subsection{Outage probability for message $x_{b}$}
In this subsection, closed-form expressions are provided for the OP of message $x_b$. These expressions are shown in Tables \ref{OPb}, \ref{OPbv2}, where all possible scenarios regarding the values of the system parameters are taken into account to provide a complete analysis of the investigated system model.
\begin{table*}
    \caption{Expressions for OP $x_{b}$ for $C_6<0$}
    \label{OPb}
    \centering
    \resizebox{\textwidth}{!}{
    \begin{tabular}{c|c|c|c|c}
        \hline \hline
        \multicolumn{4}{c|}{Conditions}
        & Expression
        \\ \hline \hline
        \multirow{19}{*}[0cm]{\rotatebox{90}{$C_7 < 0$}} &
        $C_4 < 0$ &
        &&
        1
        \\ \cline{2-5}
        &
        \multirow{8}{*}[0cm]{$C_4 \geq 0, \sqrt{C_4} < \frac{D_y}{2}$ } &
        \multirow{4}{*}[0cm]{$C_5 \geq 0, \sqrt{C_5} \leq \frac{D_y}{2}$ } &
        $\theta_2\left(4d^2+D_y^2\right)<4\eta\gamma_2$&
        $1-\Phi_3\left(\sqrt{C_4},\sqrt{C_5},\frac{D_y}{2}\right)-\frac{2\sqrt{C_5}}{D_y}$
        \\ \cline{4-5}
        &&&
        $\theta_2\left(4d^2+D_y^2\right)\geq4\eta\gamma_2,C_9\geq0,C_9\leq C_4$&
        $1-\Phi_{1}\left(\sqrt{C_9},\sqrt{C_5}\right)-\Phi_{2}\left(\sqrt{C_5},0\right)$
        \\ \cline{4-5}
        &&&
        $\theta_2\left(4d^2+D_y^2\right)\geq4\eta\gamma_2,C_9\geq0,C_9\geq C_4$&
        $1-\Phi_{1}\left(\sqrt{C_4},\sqrt{C_5}\right)-\Phi_{2}\left(\sqrt{C_5},0\right)$
        \\ \cline{4-5}
        &&&
        $\theta_2\left(4d^2+D_y^2\right)\geq4\eta\gamma_2,C_9\leq0$&
        $1-\Phi_{2}\left(\sqrt{C_5},0\right)$
        \\ \cline{3-5}
        &&
        \multirow{4}{*}[0cm]{$C_5 < 0$}&
        $\theta_2\left(4d^2+D_y^2\right)<4\eta\gamma_2$&
        $1-\Phi_3\left(\sqrt{C_4},0,\frac{D_y}{2}\right)$
        \\ \cline{4-5}
        &&&
        $\theta_2\left(4d^2+D_y^2\right)\geq4\eta\gamma_2,C_9\geq0,C_9\leq C_4$&
        $1-\Phi_1\left(\sqrt{C_9},0\right)$
        \\ \cline{4-5}
        &&&
        $\theta_2\left(4d^2+D_y^2\right)\geq4\eta\gamma_2,C_9\geq0,C_9\geq C_4$&
        $1-\Phi_1\left(\sqrt{C_4},0\right)$
        \\ \cline{4-5}
        &&&
        $\theta_2\left(4d^2+D_y^2\right)\geq4\eta\gamma_2,C_9\leq0$&
        $1$
        \\ \cline{2-5}
        &
        \multirow{10}{*}[0cm]{$\sqrt{C_4}\geq\frac{D_y}{2}$}&
        \multirow{2}{*}[0cm]{$\sqrt{C_5}\geq\frac{D_y}{2}$}&
        $\theta_2\left(4d^2+D_y^2\right)<4\eta\gamma_2$&
        $0$
        \\ \cline{4-5}
        &&&
        $\theta_2\left(4d^2+D_y^2\right)\geq4\eta\gamma_2$&
        $1-\Phi_2\left(\frac{D_y}{2},0\right)$
        \\ \cline{3-5}
        &&
        \multirow{4}{*}[0cm]{$C_5\geq0,\sqrt{C_5}\leq\frac{D_y}{2}$}&
        $\theta_2\left(4d^2+D_y^2\right)<4\eta\gamma_2$&
        $1-\Phi_3\left(\sqrt{C_9},\sqrt{C_5},\frac{D_y}{2}\right)-\frac{2\sqrt{C_5}}{D_y}$ 
        \\ \cline{4-5}
        &&&
        $\theta_2\left(4d^2+D_y^2\right)\geq4\eta\gamma_2,C_9\geq0,\sqrt{C_9}<\frac{D_y}{2}$&
        $1-\Phi_1\left(\sqrt{C_9},\sqrt{C_5}\right)-\Phi_2\left(\sqrt{C_5},0\right)$
        \\ \cline{4-5}
        &&&
        $\theta_2\left(4d^2+D_y^2\right)\geq4\eta\gamma_2,C_9\geq0,\sqrt{C_9}>\frac{D_y}{2}$&
        $1-\Phi_1\left(\frac{D_y}{2},\sqrt{C_5}\right)-\Phi_2\left(\sqrt{C_5},0\right)$
        \\ \cline{4-5}
        &&&
        $\theta_2\left(4d^2+D_y^2\right)\geq4\eta\gamma_2,C_9<0$&
        $1-\Phi_2\left(\sqrt{C_5},0\right)$
        \\ \cline{3-5}
        &&
        \multirow{4}{*}[0cm]{$C_5<0$}&
        $\theta_2\left(4d^2+D_y^2\right)<4\eta\gamma_2$&
        $1-\Phi_3\left(\frac{D_y}{2},0,\frac{D_y}{2}\right)$ 
        \\ \cline{4-5}
        &&&
        $\theta_2\left(4d^2+D_y^2\right)\geq4\eta\gamma_2,C_9\geq0,\sqrt{C_9}<\frac{D_y}{2}$&
        $1-\Phi_1\left(\sqrt{C_9},0\right)$
        \\ \cline{4-5}
        &&&
        $\theta_2\left(4d^2+D_y^2\right)\geq4\eta\gamma_2,C_9\geq0,\sqrt{C_9}\geq\frac{D_y}{2}$&
        $1-\Phi_1\left(\frac{D_y}{2},0\right)$
        \\ \cline{4-5}
        &&&
        $\theta_2\left(4d^2+D_y^2\right)\geq4\eta\gamma_2,C_9<0$&
        $1$
        \\ \hline
        \multirow{22}{*}[0cm]{\rotatebox{90}{$C_7\geq0,\sqrt{C_7}<\frac{D_y}{2}$}}&
        $C_4<0$&
        &
        &
        $1$
        \\ \cline{2-5}
        &
        \multirow{6}{*}[0cm]{$C_4\geq0,C_4< C_7$}&
        \multirow{3}{*}[0cm]{$C_5\geq0,C_5< C_7$}&
        $C_9\geq0,C_9<C_4$&
        $1-\Phi_1\left(\sqrt{C_9},\sqrt{C_5}\right)-\Phi_2\left(\sqrt{C_5},0\right)$
        \\ \cline{4-5}
        &&&
        $C_9\geq0,C_9\geq C_4$ &
        $1-\Phi_1\left(\sqrt{C_4},\sqrt{C_5}\right)-\Phi_2\left(\sqrt{C_5},0\right)$
        \\ \cline{4-5}
        &&&
        $C_9<0$&
        $1-\Phi_2\left(\sqrt{C_5},0\right)$
        \\ \cline{3-5}
        &&
        \multirow{3}{*}[0cm]{$C_5<0$}&
        $C_9\geq0,C_9<C_4$&
        $1-\Phi_1\left(\sqrt{C_9},0\right)$
        \\ \cline{4-5}
        &&&
        $C_9\geq0,C_9\geq C_4$&
        $1-\Phi_1\left(\sqrt{C_4},0\right)$
        \\ \cline{4-5}
        &&&
        $C_9<0$&
        $1$
        \\ \cline{2-5}
        &
        \multirow{7}{*}[0cm]{$C_4\geq C_7,\sqrt{C_4}\leq \frac{D_y}{2}$}&
        $C_5\geq C_7,\sqrt{C_5} < \frac{D_y}{2}$&
        &
        \tiny{$
                1\!-\!\Phi_2\!\left(\!\sqrt{C_7},\!0\!\right)\!-\!\frac{2\!\left(\!\sqrt{C_5}\!-\!\sqrt{C_7}\!\right)}{D_y}\!-\!\Phi_3\!\left(\!\sqrt{C_4},\!\sqrt{C_5},\!\frac{D_y}{2}\!\right)
                $} 
            \\ \cline{3-5}
            &&
            \multirow{3}{*}[0cm]{$C_5\geq0,C_5<C_7$}&
            $C_9\geq0,C_9<C_7$&
            \tiny{$\!1\!-\!\Phi_2\!\left(\!\sqrt{C_5},\!0\!\right)\!-\!\Phi_1\!\left(\!\sqrt{C_9},\!\sqrt{C_5}\!\right)\!-\!\Phi_3\!\left(\!\sqrt{C_4},\!\sqrt{C_7},\!\frac{D_y}{2}\!\right)$}
            \\ \cline{4-5}
            &&&
            $C_9\geq0,C_9\geq C_7$&
            \tiny{$\!1\!-\!\Phi_2\!\left(\!\sqrt{C_5},\!0\!\right)\!-\!\Phi_1\!\left(\!\sqrt{C_7},\!\sqrt{C_5}\!\right)\!-\!\Phi_3\!\left(\!\sqrt{C_4},\!\sqrt{C_7},\!\frac{D_y}{2}\!\right)$}
            \\ \cline{4-5}
            &&&
            $C_9 <0$&
            $1-\Phi_2\left(\sqrt{C_5},0\right)-\Phi_3\left(\sqrt{C_4},\sqrt{C_7},\frac{D_y}{2}\right)$
            \\ \cline{3-5}
            &&
            \multirow{3}{*}[0cm]{$C_5 <0$}&
            $C_9\geq 0,C_9<C_7$&
            $1-\Phi_1\left(\sqrt{C_9},0\right)-\Phi_3\left(\sqrt{C_4},\sqrt{C_7},\frac{D_y}{2}\right)$
            \\ \cline{4-5}
            &&&
            $C_9\geq 0,C_9\geq C_7$&
            $1-\Phi_1\left(\sqrt{C_7},0\right)-\Phi_3\left(\sqrt{C_4},\sqrt{C_7},\frac{D_y}{2}\right)$
            \\ \cline{4-5}
            &&&
            $C_9< 0$&
            $1-\Phi_3\left(\sqrt{C_4},\sqrt{C_7},\frac{D_y}{2}\right)$
            \\ \cline{2-5}
            &
            \multirow{8}{*}[0cm]{$\sqrt{C_4}\geq\frac{D_y}{2}$}&
            $\sqrt{C_5}\geq\frac{D_y}{2}$&
            &
            $1-\Phi_2\left(\sqrt{C_7},0\right)-\frac{2}{D_y}\left(\frac{D_y}{2}-\sqrt{C_7}\right)$
            \\ \cline{3-5}
            &&
            $C_5\geq C_7,\sqrt{C_5}<\frac{D_y}{2}$&
            &
            \tiny{$\!1\!-\!\Phi_2\!\left(\!\sqrt{C_7},\!0\!\right)\!-\!\frac{2\!\left(\!\sqrt{C_5}\!-\!\sqrt{C_7}\!\right)}{D_y}\!-\!\Phi_3\!\left(\!\frac{D_y}{2},\!\sqrt{C_5},\!\frac{D_y}{2}\!\right)$} 
            \\ \cline{3-5}
            &&
            \multirow{3}{*}[0cm]{$C_5\geq0,C_5<C_7$}&
            $C_9\geq0,C_9<C_7$&
            \tiny{$\!1\!-\!\Phi_2\!\left(\!\sqrt{C_5},\!0\!\right)\!-\!\Phi_1\!\left(\!\sqrt{C_9},\sqrt{C_5}\!\right)\!-\!\Phi_3\!\left(\!\frac{D_y}{2},\!\sqrt{C_7},\!\frac{D_y}{2}\!\right)$}
            \\ \cline{4-5}
            &&&
            $C_9\geq0,C_9\geq C_7$&
            \tiny{$\!1\!-\!\Phi_2\!\left(\!\sqrt{C_5},\!0\!\right)\!-\!\Phi_1\!\left(\!\sqrt{C_7},\!\sqrt{C_5}\!\right)\!-\!\Phi_3\left(\!\frac{D_y}{2},\!\sqrt{C_7},\!\frac{D_y}{2}\!\right)$} 
            \\ \cline{4-5}
            &&&
            $C_9<0$&
            $1-\Phi_2\left(\sqrt{C_5},0\right)-\Phi_3\left(\frac{D_y}{2},\sqrt{C_7},\frac{D_y}{2}\right)$
            \\ \cline{3-5}
            &&
            \multirow{3}{*}[0cm]{$C_5<0$}&
            $C_9\geq0,C_9<C_7$&
            $1-\Phi_1\left(\sqrt{C_9},0\right)-\Phi_3\left(\frac{D_y}{2},\sqrt{C_7},\frac{D_y}{2}\right)$
            \\ \cline{4-5}
            &&&
            $C_9\geq0,C_9>C_7$&
            $1-\Phi_1\left(\sqrt{C_7},0\right)-\Phi_3\left(\frac{D_y}{2},\sqrt{C_7},\frac{D_y}{2}\right)$
            \\ \cline{4-5}
            &&&
            $C_9<0$&
            $1-\Phi_3\left(\frac{D_y}{2},\sqrt{C_7},\frac{D_y}{2}\right)$
            \\ \hline 
            \multirow{14}{*}[0cm]{\rotatebox{90}{$\sqrt{C_7}\geq\frac{D_y}{2}$}}&
            $C_4<0$&
            &
            &
            1
            \\ \cline{2-5}
            &
            \multirow{6}{*}[0cm]{$C_4\geq0,\sqrt{C_4}<\frac{D_y}{2}$}&
            \multirow{3}{*}[0cm]{$C_5\geq0,\sqrt{C_5}<\frac{D_y}{2}$}&
            $C_9\geq0,C_9<C_4$&
            $1-\Phi_2\left(\sqrt{C_5},0\right)-\Phi_1\left(\sqrt{C_9},\sqrt{C_5}\right)$
            \\ \cline{4-5}
            &&&
            $C_9\geq0,C_9\geq C_4$&
            $1-\Phi_2\left(\sqrt{C_5},0\right)-\Phi_1\left(\sqrt{C_4},\sqrt{C_5}\right)$
            \\ \cline{4-5}
            &&&
            $C_9<0$&
            $1-\Phi_2\left(\sqrt{C_5},0\right)$
            \\ \cline{3-5}
            &&
            \multirow{3}{*}[0cm]{$C_5<0$}&
            $C_9\geq0,C_9<C_4$&
            $1-\Phi_1\left(\sqrt{C_9},0\right)$
            \\ \cline{4-5}
            &&&
            $C_9\geq0,C_9\geq C_4$&
            $1-\Phi_1\left(\sqrt{C_4},0\right)$
            \\ \cline{4-5}
            &&&
            $C_9<0$&
            $1$
            \\ \cline{2-5}
            &
            \multirow{7}{*}[0cm]{$\sqrt{C_4}\geq\frac{D_y}{2}$}&
            $\sqrt{C_5}\geq\frac{D_y}{2}$&
            &
            $1-\Phi_2\left(\frac{D_y}{2},0\right)$
            \\ \cline{3-5}
            &&
            \multirow{3}{*}[0cm]{$C_5\geq0,\sqrt{C_5}<\frac{D_y}{2}$}&
            $C_9\geq0,\sqrt{C_9}<\frac{D_y}{2}$&
            $1-\Phi_2\left(\sqrt{C_5},0\right)-\Phi_1\left(\sqrt{C_9},\sqrt{C_5}\right)$
            \\ \cline{4-5}
            &&&
            $C_9\geq0,\sqrt{C_9}>\frac{D_y}{2}$&
            $1-\Phi_2\left(\sqrt{C_5},0\right)-\Phi_1\left(\frac{D_y}{2},\sqrt{C_5}\right)$
            \\ \cline{4-5}
            &&&
            $C_9<0$&
            $1-\Phi_{2}\left(\sqrt{C_5},0\right)$
            \\ \cline{3-5}
            &&
            \multirow{3}{*}[0cm]{$C_5<0$}&
            $C_9\geq0,\sqrt{C_9}<\frac{D_y}{2}$&
            $1-\Phi_1\left(\sqrt{C_9},0\right)$
            \\ \cline{4-5}
            &&&
            $C_9\geq0,\sqrt{C_9}>\frac{D_y}{2}$&
            $1-\Phi_1\left(\frac{D_y}{2}\right)$
            \\ \cline{4-5}
            &&&
            $C_9<0$&
            $1$
            \\ \hline \hline
            
        \end{tabular}
        }
    \end{table*}
    \begin{table*}
        \caption{Expressions for OP $x_{b}$ for $C_6\geq0,\sqrt{C_6}<\frac{D_y}{2}$}
        \label{OPbv2}
        \centering
        \begin{tabular}{c|c|c|c|c}
            \hline \hline
            \multicolumn{4}{c|}{Conditions}
            & Expression
            \\ \hline \hline
            \multirow{14}{*}[0cm]{\rotatebox{90}{$C_7 < 0$}} &
            $C_4 < C_6$ &
            &&
            1
            \\ \cline{2-5}
            &
            \multirow{6}{*}[0cm]{$C_4\geq C_6,\sqrt{C_4}<\frac{D_y}{2}$}&
            \multirow{3}{*}[0cm]{$C_5\geq C_6,\sqrt{C_5}<\frac{D_y}{2}$}&
            $C_9\geq0,C_9<C_4$&
            $1-\Phi_2\left(\sqrt{C_5},\sqrt{C_6}\right)-\Phi_1\left(\sqrt{C_9},\sqrt{C_5}\right)$
            \\ \cline{4-5}
            &&&
            $C_9\geq0,C_9\geq C_4$&
            $1-\Phi_2\left(\sqrt{C_5},\sqrt{C_6}\right)-\Phi_1\left(\sqrt{C_4},\sqrt{C_5}\right)$
            \\ \cline{4-5}
            &&&
            $C_9<0$&
            $1-\Phi_2\left(\sqrt{C_5},\sqrt{C_6}\right)$
            \\ \cline{3-5}
            &&
            \multirow{3}{*}[0cm]{$C_5<C_6$}&
            $C_9\geq0,C_9<C_4$&
            $1-\Phi_1\left(\sqrt{C_9},\sqrt{C_6}\right)$
            \\ \cline{4-5}
            &&&
            $C_9\geq0,C_9\geq C_4$&
            $1-\Phi_1\left(\sqrt{C_4},\sqrt{C_6}\right)$
            \\ \cline{4-5}
            &&&
            $C_9<0$&
            $1$
            \\ \cline{2-5}
            &
            \multirow{7}{*}[0cm]{$\sqrt{C_4}\geq\frac{D_y}{2}$}&
            $\sqrt{C_5}\geq\frac{D_y}{2}$&
            &
            $1-\frac{2}{D_y}\left(\frac{D_y}{2}-\sqrt{C_6}\right)$
            \\ \cline{3-5}
            &&
            \multirow{3}{*}[0cm]{$C_5\geq C_6,\sqrt{C_5}<\frac{D_y}{2}$}&
            $C_9\geq0,\sqrt{C_9}<\frac{D_y}{2}$&
            $1-\frac{2}{D_y}\left(\sqrt{C_5}-\sqrt{C_6}\right)-\Phi_1\left(\sqrt{C_9},\sqrt{C_5}\right)$
            \\ \cline{4-5}
            &&&
            $C_9\geq0,\sqrt{C_9}\geq\frac{D_y}{2}$&
            $1-\frac{2}{D_y}(\sqrt{C_5}-\sqrt{C_6})-\Phi_1\left(\frac{D_y}{2},\sqrt{C_5}\right)$
            \\ \cline{4-5}
            &&&
            $C_9<0$&
            $1-\frac{2}{D_y}(\sqrt{C_5}-\sqrt{C_6})$
            \\ \cline{3-5}
            &&
            \multirow{3}{*}[0cm]{$C_5<C_6$}&
            $C_9\geq0,\sqrt{C_9}<\frac{D_y}{2}$&
            $1-\Phi_3\left(\sqrt{C_9},\sqrt{C_6},\frac{D_y}{2}\right)$
            \\ \cline{4-5}
            &&&
            $C_9\geq0,\sqrt{C_9}\geq\frac{D_y}{2}$&
            $1-\Phi_3\left(\frac{D_y}{2},\sqrt{C_6},\frac{D_y}{2}\right)$
            \\ \cline{4-5}
            &&&
            $C_9<0$&
            $1$
            \\ \hline
            \multirow{22}{*}[0cm]{\rotatebox{90}{$C_7^2\geq0,C_7<\frac{D_y}{2}$}}&
            $C_4<C_6$&
            &
            &
            $1$
            \\ \cline{2-5}
            &
            \multirow{6}{*}[0cm]{$C_4\geq C_6,C_4<C_7$}&
            \multirow{3}{*}[0cm]{$C_5\geq C_6,C_5<C_7$}&
            $C_9\geq0,C_9<C_4$&
            $1-\Phi_2\left(\sqrt{C_5},\sqrt{C_6}\right)-\Phi_1\left(\sqrt{C_9},\sqrt{C_5}\right)$
            \\ \cline{4-5}
            &&&
            $C_9\geq0,C_9\geq C_4$&
            $1-\Phi_2\left(\sqrt{C_5},\sqrt{C_6}\right)-\Phi_1\left(\sqrt{C_4},\sqrt{C_5}\right)$
            \\ \cline{4-5}
            &&&
            $C_9<0$&
            $1-\Phi_2\left(\sqrt{C_5},\sqrt{C_6}\right)$
            \\ \cline{3-5}
            &&
            $C_5<C_6$&
            $C_9\geq0,C_9<C_4$&
            $1-\Phi_1\left(\sqrt{C_9},\sqrt{C_6}\right)$
            \\ \cline{4-5}
            &&&
            $C_9\geq0,C_9\geq C_4$&
            $1-\Phi_1\left(\sqrt{C_4},\sqrt{C_6}\right)$
            \\ \cline{4-5}
            &&&
            $C_9<0$&
            $1$
            \\ \cline{2-5}
            &
            \multirow{7}{*}[0cm]{$C_4\geq C_7,\sqrt{C_4}<\frac{D_y}{2}$}&
            $C_5\geq C_7,\sqrt{C_5}<\frac{D_y}{2}$&
            &
            \tiny{$\!1\!-\!\Phi_2\!\left(\! \sqrt{C_7},\! \sqrt{C_6}\!\right)\!-\!\frac{2\!(\! \sqrt{C_5}\!-\! \sqrt{C_7}\!)}{D_y}\!-\!\Phi_3\!\left(\!\sqrt{ C_4},\! \sqrt{C_5},\!\frac{D_y}{2}\!\right)$}
            \\ \cline{3-5}
            &&
            \multirow{3}{*}[0cm]{$C_5\geq C_6,C_5<C_7$}&
            $C_9\geq0,C_9<C_7$&
            \tiny{$\!1\!-\!\Phi_2\!\left(\! \sqrt{C_5},\! \sqrt{C_6}\!\right)\!-\!\Phi_1\!\left(\! \sqrt{C_9},\! \sqrt{C_5}\!\right)\!-\!\Phi_3\!\left(\! \sqrt{C_4},\! \sqrt{C_7},\!\frac{D_y}{2}\!\right)$}
            \\ \cline{4-5}
            &&&
            $C_9\geq0,C_9\geq C_7$&
            \tiny{$\!1\!-\!\Phi_2\!\left(\! \sqrt{C_5},\! \sqrt{C_6}\!\right)\!-\!\Phi_1\!\left(\! \sqrt{C_7},\! \sqrt{C_5}\!\right)\!-\!\Phi_3\!\left(\! \sqrt{C_4},\! \sqrt{C_7},\!\frac{D_y}{2}\!\right)$}
            \\ \cline{4-5}
            &&&
            $C_9<0$&
            $1-\Phi_2\left(\sqrt{C_5},\sqrt{C_6}\right)-\Phi_3\left(\sqrt{C_4},\sqrt{C_7},\frac{D_y}{2}\right)$
            \\ \cline{3-5}
            &&
            $C_5<C_6$&
            $C_9\geq0,C_9<C_7$&
            $1-\Phi_1\left(\sqrt{C_9},\sqrt{C_6}\right)-\Phi_3\left(\sqrt{C_4},\sqrt{C_7},\frac{D_y}{2}\right)$
            \\ \cline{4-5}
            &&&
            $C_9\geq0,C_9\geq C_7$&
            $1-\Phi_1\left(\sqrt{C_7},\sqrt{C_6}\right)-\Phi_3\left(\sqrt{C_4},\sqrt{C_7},\frac{D_y}{2}\right)$
            \\ \cline{4-5}
            &&&
            $C_9<0$&
            $1-\Phi_3\left(\sqrt{C_4},\sqrt{C_7},\frac{D_y}{2}\right)$
            \\ \cline{2-5}
            &
            \multirow{8}{*}[0cm]{$\sqrt{C_4}\geq\frac{D_y}{2}$}&
            $\sqrt{C_5}\geq\frac{D_y}{2}$&
            &
            $1-\Phi_2\left(\sqrt{C_7},\sqrt{C_6}\right)-\frac{2}{D_y}\left(\frac{D_Y}{2}-\sqrt{C_7}\right)$
            \\ \cline{3-5}
            &&
            $C_5\geq C_7,\sqrt{C_5}<\frac{D_y}{2}$&
            &
            \tiny{$\!1\!-\!\Phi_2\!\left(\! \sqrt{C_7},\! \sqrt{C_6}\!\right)\!-\!\frac{2\!(\! \sqrt{C_5}\!-\!\sqrt{C_7}\!)}{D_y}\!-\!\Phi_3\!\left(\!\frac{D_y}{2},\! \sqrt{C_5},\!\frac{D_y}{2}\!\right)$}
            \\ \cline{3-5}
            &&
            \multirow{3}{*}[0cm]{$C_5\geq C_6,C_5<C_7$}&
            $C_9\geq0,C_9<C_7$&
            \tiny{$\!1-\!\Phi_2\!\left(\! \sqrt{C_5},\! \sqrt{C_6}\!\right)\!-\!\Phi_1\!\left(\! \sqrt{C_9},\! \sqrt{C_5}\!\right)\!-\!\Phi_3\!\left(\!\frac{D_y}{2},\! \sqrt{C_7},\!\frac{D_y}{2}\right)$}
            \\ \cline{4-5}
            &&&
            $C_9\geq0,C_9\geq C_7$&
            \tiny{$\!1-\!\Phi_2\!\left(\! \sqrt{C_5},\! \sqrt{C_6}\!\right)\!-\!\Phi_1\!\left(\! \sqrt{C_7},\! \sqrt{C_5}\!\right)\!-\!\Phi_3\!\left(\!\frac{D_y}{2},\! \sqrt{C_7},\!\frac{D_y}{2}\right)$}
            \\ \cline{4-5}
            &&&
            $C_9<0$&
            $1-\Phi_2\left(\sqrt{C_5},\sqrt{C_6}\right)-\Phi_3\left(\frac{D_y}{2},\sqrt{C_7},\frac{D_y}{2}\right)$
            \\ \cline{3-5}
            &&
            \multirow{3}{*}[0cm]{$C_5<C_6$}&
            $C_9\geq0,C_9<C_7$&
            $1-\Phi_1\left(\sqrt{C_9},\sqrt{C_6}\right)-\Phi_3\left(\frac{D_y}{2},\sqrt{C_7},\frac{D_y}{2}\right)$
            \\ \cline{4-5}
            &&&
            $C_9\geq0,C_9\geq C_7$&
            $1-\Phi_1\left(\sqrt{C_7},\sqrt{C_6}\right)-\Phi_3\left(\frac{D_y}{2},\sqrt{C_7},\frac{D_y}{2}\right)$
            \\ \cline{4-5}
            &&&
            $C_9<0$&
            $1-\Phi_3\left(\frac{D_y}{2},\sqrt{C_7},\frac{D_y}{2}\right)$
            \\ \hline
            \multirow{14}{*}[0cm]{\rotatebox{90}{$\sqrt{C_7}\geq\frac{D_y}{2}$}}&
            $C_4<C_6$&
            &
            &
            $1$
            \\ \cline{2-5}
            &
            \multirow{6}{*}[0cm]{$C_4\geq C_6,\sqrt{C_4}<\frac{D_y}{2}$}&
            \multirow{3}{*}[0cm]{$C_5\geq C_6,\sqrt{C_5}<\frac{D_y}{2}$}&
            $C_9\geq0,C_9<C_4$&
            $1-\Phi_2\left(\sqrt{C_5},\sqrt{C_6}\right)-\Phi_1\left(\sqrt{C_9},\sqrt{C_5}\right)$
            \\ \cline{4-5}
            &&&
            $C_9\geq0,C_9\geq C_4$&
            $1-\Phi_2\left(\sqrt{C_5},\sqrt{C_6}\right)-\Phi_1\left(\sqrt{C_4},\sqrt{C_5}\right)$
            \\ \cline{4-5}
            &&&
            $C_9<0$&
            $1-\Phi_2\left(\sqrt{C_5},\sqrt{C_6}\right)$
            \\ \cline{3-5}
            &&
            \multirow{3}{*}[0cm]{$C_5<C_6$}&
            $C_9\geq0,C_9<C_4$&
            $1-\Phi_1\left(\sqrt{C_9},\sqrt{C_6}\right)$
            \\ \cline{4-5}
            &&&
            $C_9\geq0,C_9\geq C_4$&
            $1-\Phi_1\left(\sqrt{C_4},\sqrt{C_6}\right)$
            \\ \cline{4-5}
            &&&
            $C_9<0$&
            $1$
            \\ \cline{2-5}
            &
            \multirow{7}{*}[0cm]{$\sqrt{C_4}>\frac{D_y}{2}$}&
            $\sqrt{C_5}\geq\frac{D_y}{2}$&
            &
            $1-\Phi_2\left(\frac{D_y}{2},\sqrt{C_6}\right)$
            \\ \cline{3-5}
            &&
            \multirow{3}{*}[0cm]{$C_5\geq C_6,\sqrt{C_5}<\frac{D_y}{2}$}&
            $C_9\geq0,\sqrt{C_9}<\frac{D_y}{2}$&
            $1-\Phi_2\left(\sqrt{C_5},\sqrt{C_6}\right)-\Phi_1\left(\sqrt{C_9},\sqrt{C_5}\right)$
            \\ \cline{4-5}
            &&&
            $C_9\geq0,\sqrt{C_9}\geq\frac{D_y}{2}$&
            $1-\Phi_2\left(\sqrt{C_5},\sqrt{C_6}\right)-\Phi_1\left(\frac{D_y}{2},\sqrt{C_5}\right)$
            \\ \cline{4-5}
            &&&
            $C_9<0$&
            $1-\Phi_2\left(\sqrt{C_5},\sqrt{C_6}\right)$
            \\ \cline{3-5}
            &&
            \multirow{3}{*}[0cm]{$C_5<C_6$}&
            $C_9\geq0,\sqrt{C_9}<\frac{D_y}{2}$&
            $1-\Phi_1\left(\sqrt{C_9},\sqrt{C_6}\right)$
            \\ \cline{4-5}
            &&&
            $C_9\geq0,\sqrt{C_9}\geq\frac{D_y}{2}$&
            $1-\Phi_1\left(\frac{D_y}{2},\sqrt{C_6}\right)$
            \\ \cline{4-5}
            &&&
            $C_9<0$&
            $1$
            \\ \hline
        \end{tabular}
    \end{table*}
\begin{IEEEproof} \label{ProofXb}
    Since SIC is implemented in the proposed scheme, successfully decoding message $x_{1a}$ is required before attempting to decode $x_{b}$. Thus, to calculate the OP for $x_{b}$, the OP of $x_{1a}$ must be taken into account. This can be achieved by considering the  event of successfully decoding $x_{b}$ and $x_{1a}$, and then taking the complementary event. The received messages are successfully decoded when the  SINR is greater than the corresponding threshold. Starting from
    \begin{equation} \small
            P_{s,x_{b}} = \Pr\left(\underbrace{\gamma_b\geq \theta_2}_{E_1},\underbrace{\gamma_{1a}\geq \theta_{11}}_{E_2}\right)
    \end{equation}
and after some algebraic manipulations we derive that
\begin{equation} \small
    P_1 = \Pr\left(y_{U,2}^2+d^2\leq \frac{\eta\gamma_2\left(y_{U,1}^2+d^2\right)}{\theta_2\left((1-\alpha)\eta\gamma_1+y_{U,1}^2+d^2\right)}\right),
\end{equation}
where $P_1 = \Pr\left(E_1\right)$ and
\begin{equation} \small
    P_2 \!=\! \Pr\!\left(\! y_{U,2}^2\!+\! d^2\!\geq\!\frac{\eta\gamma_2\theta_{11}\left(y_{U,1}^2+d^2\right)}{\eta\gamma_1\!\left(\! (\alpha\!-\!(1\!-\!\alpha)\theta_{11})\!-\!\theta_{11}\!\left(y_{U,1}^2+d^2\right)\!\right)}\right),
\end{equation} 
where $P_2 = \Pr\left(E_2\right)$. 
Taking these into account, we have that $C_1\leq y_{U,2}^2 \leq C_2$, where $C_2 = \frac{\eta\gamma_2\left(y_{U,1}^2+d^2\right)}{\theta_2\left((1-\alpha)\eta\gamma_1+y_{U,1}^2+d^2\right)}-d^2$. For $C_1<C_2$ to be true, $y_{U,1}<\sqrt{C_9}$ must hold. It can be easily derived that when $y_{U,1}\geq \sqrt{C_5}$ $C_1$ is positive and $y_{U,1}\geq \sqrt{C_4}$, thus $\sqrt{C_1} \geq \frac{D_y}{2}$. Regarding $C_2$, when $y_{U,1} \geq \sqrt{C_6}$ $C_2$ is positive, while when $y_{U,1}\geq \sqrt{C_7}$, $C_2$ is greater than $\frac{D_y^2}{4}$. Considering all possible values for each term the numerous cases presented in Tables \ref{OPb}, \ref{OPbv2} can be calculated. Indicatively, the way to calculate the final expressions for one of the cases is presented below. Assuming $C_6<0$, $C_7\geq0$ and $\sqrt{C_7}<\frac{D_y}{2}$, based on the previous expressions it is derived that $\sqrt{C_2}>0$ since $y_{U,1}\geq \sqrt{C_6}$, for $y_{U,1}\geq \sqrt{C_7}$ the upper limit of the inner integral must be $\frac{D_y}{2}$, while for $y_{U,1}<\sqrt{C_7}$ the upper limit must be $\sqrt{C_2}$. Furthermore, assuming $\sqrt{C_4}\geq \sqrt{C_7}$, $\sqrt{C_4}<\frac{D_y}{2}$ and $C_5\geq0$, $\sqrt{C_5}<\sqrt{C_7}$ and taking advantage of the symmetry of the system model, the probability of successfully decoding $x_b$ is derived as
\begin{equation} \small
\begin{aligned}
P_{s,x_b} = & 4\underbrace{\int_{0}^{\sqrt{C_5}}\!\!\!\int_{0}^{\sqrt{C_2}}\!\!\!\frac{1}{D_y^2}\mathrm{d} y_{U,2}\mathrm{d} y_{U,1}}_{I_3} \!+4\!\underbrace{\int_{\sqrt{C_5}}^{\sqrt{C_7}}\!\!\!\int_{\sqrt{C_1}}^{\sqrt{C_2}}\!\frac{1}{D_y^2}\mathrm{d} y_{U,2}\mathrm{d} y_{U,1}}_{I_4} \\
&+4\underbrace{\int_{\sqrt{C_7}}^{\sqrt{C_4}}\int_{\sqrt{C_1}}^{\frac{D_y}{2}}\frac{1}{D_y^2}\mathrm{d} y_{U,2}\mathrm{d} y_{U,1}}_{I_5}.
\end{aligned}
\end{equation}
Regarding the first integral, it assumes the form of
\begin{equation} \small
I_3 \!=\! \frac{4}{D_y^2}\int_{0}^{\sqrt{C_5}}\sqrt{\frac{\eta\gamma_2\left(y_{U,1}^2+d^2\right)}{\theta_2\left((1-\alpha)\eta\gamma_1+y_{U,1}^2+d^2\right)}-d^2}\mathrm{d} y_{U,1},
\end{equation}
which after some algebraic manipulations can be written as
\begin{equation} \small
\label{I3}
I_3=\frac{4}{D_y^2}\int_{0}^{C_5}\sqrt{\frac{Q_2y_{U,1}^2+P_2}{S_2y_{U,1}^2+R_2}}\mathrm{d} y_{U,1},
\end{equation}
where $Q_2,S_2,S_2$, and $T_2$ are provided in Table \ref{HelpingFunctions}. To derive a closed-form expression for \eqref{I3}, \cite[(3.169.2)]{gradshteyn2014table} is used. Specifically, it can be calculated that $I_3 = \Phi_2\left(C_5,0\right)$, where $\Phi_2$ is given in \eqref{Phi2}. 
Moving on, since the limits of the inner integral of $I_4$ is $\sqrt{C_1}$ and $\sqrt{C_2}$, it must be ensured that $\sqrt{C_1}<\sqrt{C_2}$, thus a comparison between $\sqrt{C_7}$ and $\sqrt{C_9}$ must be made. Assuming $\sqrt{C_7}<\sqrt{C_9}$, this expression holds, while if $\sqrt{C_9}<\sqrt{C_7}$ $\sqrt{C_9}$ must be used as the upper limit. Finally, if $C_9<0$, which means that $y_{U,1}\geq \sqrt{C_9}$ and $\sqrt{C_2}<\sqrt{C_1}$, $I_4=0$. As previously, this integral can be written as
\begin{equation} \small
\begin{aligned}
I_4 =& \frac{4}{D_y^2}\int_{\sqrt{C_5}}^{\sqrt{C_7}}\sqrt{\frac{Q_2y_{U,1}^2+P_2}{S_2y_{U,1}^2+R_2}}\mathrm{d} y_{U,1} \\ &-\frac{4}{D_y^2}\int_{\sqrt{C_5}}^{\sqrt{C_7}}\sqrt{\frac{Q_1y_{U,1}^2+P_1}{R_1-S_1y_{U,1}^2}}\mathrm{d} y_{U,1}.
\end{aligned}
\end{equation} 
These integrals have already been calculated, thus the analytical procedure is omitted. It can be derived that $I_4 = \Phi_1\left(\sqrt{C_7},\sqrt{C_5}\right)$. Finally, $I_5$ assumes the form of 
\begin{equation} \small
I_5 = \frac{4}{D_y^2}\int_{\sqrt{C_7}}^{\sqrt{C_4}}\frac{D_y}{2} \mathrm{d} y_{U,1}-\frac{4}{D_y^2}\int_{\sqrt{C_7}}^{\sqrt{C_4}}\sqrt{\frac{Q_1y_{U,1}^2+P_1}{R_1-S_1y_{U,1}^2}}\mathrm{d} y_{U,1},
\end{equation}
which can be calculated as $I_5 = \Phi_3\left(\sqrt{C_7},\sqrt{C_4},\frac{D_y}{2}\right)$. To calculate the OP of $x_b$, the complementary event must be taken, thus $P_{o,x_b} = 1 - P_{s,x_b}$ from which the final expressions are calculated. Following a similar procedure, the rest of the cases can also be derived. 
\end{IEEEproof}
\subsection{Outage probability for message $x_{2a}$}
In this subsection, closed-form expressions for the OP of message $x_{2a}$ are derived. These expressions are provided in Tables 
\ref{OP2av1}-\ref{OP2av6}.

    \begin{table*}
        \caption{Expressions for OP $x_{2a}$ $C_6<0$}
        \label{OP2av1}
        \centering
        \begin{tabular}{c|c|c|c|c|c}
            \hline \hline
            \multicolumn{5}{c|}{Conditions}
            & Expression
            \\ \hline \hline
            \multirow{39}{*}[0cm]{\rotatebox{90}{$C_7 < 0$}} &
            $C_8<0$&
            &
            &
            &
            $1$
            \\ \cline{2-6}
            &
            \multirow{19}{*}[0cm]{\rotatebox{90}{$C_8 \geq 0,\sqrt{C_8}<\frac{D_y}{2}$}} &
            $C_4<0$&
            &
            &
            1
            \\ \cline{3-6}
            &&
            \multirow{8}{*}[0cm]{\rotatebox{90}{$C_4 \geq 0,C_4<C_8$}} &
            \multirow{4}{*}[0cm]{$C_5\geq0,C_5<C_8$}&
            $\theta_2\left(4d^2+D_y^2\right)<4\eta\gamma_2$&
            $1-\Phi_3\left(\sqrt{C_4},\sqrt{C_5},\frac{D_y}{2}\right)-\frac{2\sqrt{C_5}}{D_y}$
            \\ \cline{5-6}
            &&&&
            $\theta_2\left(4d^2+D_y^2\right)\geq4\eta\gamma_2,C_9\geq0,C_9<C_4$&
            $1-\Phi_1\left(\sqrt{C_9},\sqrt{C_5}\right)-\Phi_2\left(\sqrt{C_5},0\right)$
            \\ \cline{5-6}
            &&&&
            $\theta_2\left(4d^2+D_y^2\right)\geq4\eta\gamma_2,C_9\geq0,C_9\geq C_4$&
            $1-\Phi_1\left(\sqrt{C_4},\sqrt{C_5}\right)-\Phi_2\left(\sqrt{C_5},0\right)$
            \\ \cline{5-6}
            &&&&
            $\theta_2\left(4d^2+D_y^2\right)\geq4\eta\gamma_2,C_9<0$&
            $1-\Phi_2\left(\sqrt{C_5},0\right)$
            \\ \cline{4-6}
            &&&
            \multirow{4}{*}[0cm]{$C_5<0$}&
            $\theta_2\left(4d^2+D_y^2\right)<4\eta\gamma_2$&
            $1-\Phi_3\left(\sqrt{C_4},0,\frac{D_y}{2}\right)$
            \\ \cline{5-6}
            &&&&
            $\theta_2\left(4d^2+D_y^2\right)\geq4\eta\gamma_2,C_9\geq0,C_9<C_4$&
            $1-\Phi_1\left(\sqrt{C_9},0\right)$
            \\ \cline{5-6}
            &&&&
            $\theta_2\left(4d^2+D_y^2\right)\geq4\eta\gamma_2,C_9\geq0,C_9\geq C_4$&
            $1-\Phi_1\left(\sqrt{C_4},0\right)$
            \\ \cline{5-6}
            &&&&
            $\theta_2\left(4d^2+D_y^2\right)\geq4\eta\gamma_2,C_9<0$&
            $1$
            \\ \cline{3-6}
            &&
            \multirow{10}{*}[0cm]{\rotatebox{90}{$C_4 \geq C_8$}} &
            \multirow{2}{*}[0cm]{$C_5\geq C_8$}&
            $\theta_2\left(4d^2+D_y^2\right)<4\eta\gamma_2$&
            $1-\frac{2\sqrt{C_8}}{D_y}$
            \\ \cline{5-6}
            &&&&
            $\theta_2\left(4d^2+D_y^2\right)\geq4\eta\gamma_2$&
            $1-\Phi_2\left(\sqrt{C_8},0\right)$
            \\ \cline{4-6}
            &&&
            \multirow{4}{*}[0cm]{$C_5\geq0,C_5< C_8$}&
            $\theta_2\left(4d^2+D_y^2\right)<4\eta\gamma_2$&
            $1-\frac{2\sqrt{C_5}}{D_y}-\Phi_3\left(\sqrt{C_8},\sqrt{C_5},\frac{D_y}{2}\right)$
            \\ \cline{5-6}
            &&&&
            $\theta_2\left(4d^2+D_y^2\right)\geq4\eta\gamma_2,C_9\geq0,C_9<C_8$&
            $1-\Phi_2\left(\sqrt{C_5},0\right)-\Phi_1\left(\sqrt{C_9},\sqrt{C_5}\right)$
            \\ \cline{5-6}
            &&&&
            $\theta_2\left(4d^2+D_y^2\right)\geq4\eta\gamma_2,C_9\geq0,C_9\geq C_8$&
            $1-\Phi_2\left(\sqrt{C_5},0\right)-\Phi_1\left(\sqrt{C_8},\sqrt{C_5}\right)$
            \\ \cline{5-6}
            &&&&
            $\theta_2\left(4d^2+D_y^2\right)\geq4\eta\gamma_2,C_9<0$&
            $1-\Phi_2\left(\sqrt{C_5},0\right)$
            \\ \cline{4-6}
            &&&
            \multirow{3}{*}[0cm]{$C_5<0$}&
            $\theta_2\left(4d^2+D_y^2\right)<4\eta\gamma_2$&
            $1-\Phi_3\left(\sqrt{C_8},0,\frac{D_y}{2}\right)$
            \\ \cline{5-6}
            &&&&
            $\theta_2\left(4d^2+D_y^2\right)\geq4\eta\gamma_2,C_9\geq0,C_9<C_8$&
            $1-\Phi_1\left(\sqrt{C_9},0\right)$
            \\ \cline{5-6}
            &&&&
            $\theta_2\left(4d^2+D_y^2\right)\geq4\eta\gamma_2,C_9\geq0,C_9\geq C_8$&
            $1-\Phi_1\left(\sqrt{C_8},0\right)$
            \\ \cline{5-6}
            &&&&
            $\theta_2\left(4d^2+D_y^2\right)\geq4\eta\gamma_2,C_9<0$&
            $1$
            \\ \cline{2-6}
            &
            \multirow{19}{*}[0cm]{\rotatebox{90}{$\sqrt{C_8}\geq\frac{D_y}{2}$}}& 
            $C_4<0$&
            &&
            $1$
            \\ \cline{3-6}
            &&
            \multirow{8}{*}[0cm]{\rotatebox{90}{$C_4\geq0,\sqrt{C_4}<\frac{D_y}{2}$}}&
            \multirow{4}{*}[0cm]{$C_5\geq0,\sqrt{C_5}<\frac{D_y}{2}$}&
            $\theta_2\left(4d^2+D_y^2\right)<4\eta\gamma_2$&
            $1-\frac{2\sqrt{C_5}}{D_y}-\Phi_3\left(\sqrt{C_4},\sqrt{C_5},\frac{D_y}{2}\right)$
            \\ \cline{5-6}
            &&&&
            $\theta_2\left(4d^2+D_y^2\right)\geq4\eta\gamma_2,C_9\geq0,C_9<C_4$&
            $1-\Phi_2\left(\sqrt{C_5},0\right)-\Phi_1\left(\sqrt{C_9},\sqrt{C_5}\right)$
            \\ \cline{5-6}
            &&&&
            $\theta_2\left(4d^2+D_y^2\right)\geq4\eta\gamma_2,C_9\geq0,C_9\geq C_4$&
            $1-\Phi_2\left(\sqrt{C_5},0\right)-\Phi_1\left(\sqrt{C_4},\sqrt{C_5}\right)$
            \\ \cline{5-6}
            &&&&
            $\theta_2\left(4d^2+D_y^2\right)\geq4\eta\gamma_2,C_9<0$&
            $1-\Phi_2\left(\sqrt{C_5},0\right)$
            \\ \cline{4-6}
            &&&
            \multirow{4}{*}[0cm]{$C_5<0$}&
            $\theta_2\left(4d^2+D_y^2\right)<4\eta\gamma_2$&
            $1-\Phi_3\left(\sqrt{C_4},0,\frac{D_y}{2}\right)$
            \\ \cline{5-6}
            &&&&
            $\theta_2\left(4d^2+D_y^2\right)\geq4\eta\gamma_2,C_9\geq0,C_9<C_4$&
            $1-\Phi_1\left(\sqrt{C_9},0\right)$
            \\ \cline{5-6}
            &&&&
            $\theta_2\left(4d^2+D_y^2\right)\geq4\eta\gamma_2,C_9\geq0,C_9\geq C_4$&
            $1-\Phi_1\left(\sqrt{C_4},0\right)$
            \\ \cline{5-6}
            &&&&
            $\theta_2\left(4d^2+D_y^2\right)\geq4\eta\gamma_2,C_9<0$&
            $1$
            \\ \cline{3-6}
            &&
            \multirow{10}{*}[0cm]{\rotatebox{90}{$\sqrt{C_4}\geq\frac{D_y}{2}$}}&
            \multirow{2}{*}[0cm]{$\sqrt{C_5}\geq\frac{D_y}{2}$}&
            $\theta_2\left(4d^2+D_y^2\right)<4\eta\gamma_2$&
            $0$
            \\ \cline{5-6}
            &&&&
            $\theta_2\left(4d^2+D_y^2\right)\geq4\eta\gamma_2$&
            $1-\Phi_2\left(\frac{D_y}{2},0\right)$
            \\ \cline{4-6}
            &&&
            \multirow{4}{*}[0cm]{$C_5\geq0,\sqrt{C_5}<\frac{D_y}{2}$}&
            $\theta_2\left(4d^2+D_y^2\right)<4\eta\gamma_2$&
            $1-\frac{2\sqrt{C_5}}{D_y}-\Phi_3\left(\frac{D_y}{2},\sqrt{C_5},\frac{D_y}{2}\right)$
            \\ \cline{5-6}
            &&&&
            $\theta_2\left(4d^2+D_y^2\right)\geq4\eta\gamma_2,C_9\geq0,C_9<\frac{D_y}{2}$&
            $1-\Phi_2\left(\sqrt{C_5},0\right)-\Phi_1\left(\sqrt{C_9},\sqrt{C_5}\right)$
            \\ \cline{5-6}
            &&&&
            $\theta_2\left(4d^2+D_y^2\right)\geq4\eta\gamma_2,C_9\geq0,C_9\geq\frac{D_y}{2}$&
            $1-\Phi_2\left(\sqrt{C_5},0\right)-\Phi_1\left(\frac{D_y}{2},\sqrt{C_5}\right)$
            \\ \cline{5-6}
            &&&&
            $\theta_2\left(4d^2+D_y^2\right)\geq4\eta\gamma_2,C_9<0,$&
            $1-\Phi_2\left(\sqrt{C_5},0\right)$
            \\ \cline{4-6}
            &&&
            \multirow{4}{*}[0cm]{$C_5<0$}&
            $\theta_2\left(4d^2+D_y^2\right)<4\eta\gamma_2$&
            $1-\Phi_3\left(\frac{D_y}{2},0,\frac{D_y}{2}\right)$
            \\ \cline{5-6}
            &&&&
            $\theta_2\left(4d^2+D_y^2\right)\geq4\eta\gamma_2,C_9\geq0,\sqrt{C_9}<\frac{D_y}{2}$&
            $1-\Phi_1\left(\sqrt{C_9},0\right)$
            \\ \cline{5-6}
            &&&&
            $\theta_2\left(4d^2+D_y^2\right)\geq4\eta\gamma_2,C_9\geq0,\sqrt{C_9}\geq\frac{D_y}{2}$&
            $1-\Phi_1\left(\frac{D_y}{2},0\right)$
            \\ \cline{5-6}
            &&&&
            $\theta_2\left(4d^2+D_y^2\right)\geq4\eta\gamma_2,C_9<0$&
            $1$
            \\ \hline \hline
        \end{tabular}
    \end{table*}
    \begin{table*}
        \caption{Expressions for OP $x_{2a}$ $C_6<0$ v2}
        \label{OP2av2}
        \centering
        \begin{tabular}{c|c|c|c|c|c}
            \hline \hline
            \multicolumn{5}{c|}{Conditions}
            & Expression
            \\ \hline \hline
            \multirow{59}{*}[0cm]{\rotatebox{90}{$C_7 \geq 0, \sqrt{C_7} <\frac{D_y}{2}$}} &
            $C_8<0$&
            &
            &
            &
            $1$
            \\ \cline{2-6}
            &
            \multirow{14}{*}[0cm]{\rotatebox{90}{$C_8\geq0,C_8<C_7$}}&
            $C_4<0$&
            &
            &
            1
            \\ \cline{3-6}
            &&
            \multirow{6}{*}[0cm]{\rotatebox{90}{$\! C_4 \! \geq \!0\!,\! C_4\!<\! C_8$}}&
            \multirow{3}{*}[0cm]{$C_5\geq0,C_5<C_8$}&
            $C_9\geq0,C_9<C_4$&
            $1-\Phi_1\left(\sqrt{C_9},\sqrt{C_5}\right)-\Phi_2\left(\sqrt{C_5},0\right)$
            \\ \cline{5-6}
            &&&&
            $C_9\geq0,C_9\geq C_4$&
            $1-\Phi_1\left(\sqrt{C_4},\sqrt{C_5}\right)-\Phi_2\left(\sqrt{C_5},0\right)$
            \\ \cline{5-6}
            &&&&
            $C_9<0$&
            $1-\Phi_2\left(\sqrt{C_5},0\right)$
            \\ \cline{4-6}
            &&&
            \multirow{3}{*}[0cm]{$C_5<0$}&
            $C_9\geq0,C_9<C_4$&
            $1-\Phi_1\left(\sqrt{C_9},0\right)$
            \\ \cline{5-6}
            &&&&
            $C_9\geq0,C_9\geq C_4$&
            $1-\Phi_1\left(\sqrt{C_4},0\right)$
            \\ \cline{5-6}
            &&&&
            $C_9<0$&
            $1$
            \\ \cline{3-6}
            &&
            \multirow{7}{*}[0cm]{\rotatebox{90}{$\! C_4\!\geq\! C_8\!,\! \sqrt{C_4}\!<\!\frac{D_y}{2}$}}&
            $C_5\geq C_8,\sqrt{C_5}<\frac{D_y}{2}$&
            &
            $1-\Phi_2\left(\sqrt{C_8},0\right)$
            \\ \cline{4-6}
            &&&
            \multirow{3}{*}[0cm]{$C_5\geq0,C_5<C_8$}&
            $C_9\geq0,C_9<C_8$&
            $1-\Phi_2\left(\sqrt{C_5},0\right)-\Phi_1\left(\sqrt{C_9},\sqrt{C_5}\right)$
            \\ \cline{5-6}
            &&&&
            $C_9\geq0,C_9\geq C_8$&
            $1-\Phi_2\left(\sqrt{C_5},0\right)-\Phi_1\left(\sqrt{C_8},\sqrt{C_5}\right)$
            \\ \cline{5-6}
            &&&&
            $C_9<0$&
            $1-\Phi_2\left(\sqrt{C_5},0\right)$
            \\ \cline{4-6}
            &&&
            \multirow{3}{*}[0cm]{$C_5<0$}&
            $C_9\geq0,C_9<C_8$&
            $1-\Phi_1\left(\sqrt{C_9},0\right)$
            \\ \cline{5-6}
            &&&&
            $C_9\geq0,C_9\geq C_8$&
            $1-\Phi_1\left(\sqrt{C_8},0\right)$
            \\ \cline{5-6}
            &&&&
            $C_9<0$&
            $1$
            \\ \cline{2-6}
            &
            \multirow{22}{*}[0cm]{\rotatebox{90}{$C_8\geq C_7,\sqrt{C_8}<\frac{D_y}{2}$}}&
            $C_4<0$&
            &&
            $1$
            \\ \cline{3-6}
            &&
            \multirow{6}{*}{\rotatebox{90}{$\! C_4\!\geq\!0\!,\! C_4\!<\! C_7$}}&
            \multirow{3}{*}[0cm]{$C_5\geq0,C_5<C_7$}&
            $C_9\geq0,C_9<C_4$&
            $1-\Phi_2\left(\sqrt{C_5},0\right)-\Phi_1\left(\sqrt{C_9},\sqrt{C_5}\right)$
            \\ \cline{5-6}
            &&&&
            $C_9\geq0,C_9\geq C_4$&
            $1-\Phi_2\left(\sqrt{C_5},0\right)-\Phi_1\left(\sqrt{C_4},\sqrt{C_5}\right)$
            \\ \cline{5-6}
            &&&&
            $C_9<0$&
            $1-\Phi_2\left(\sqrt{C_5},0\right)$
            \\ \cline{4-6}
            &&&
            \multirow{3}{*}[0cm]{$C_5<0$}&
            $C_9\geq0,C_9<C_4$&
            $1-\Phi_1\left(\sqrt{C_9},0\right)$
            \\ \cline{5-6}
            &&&&			
            $C_9\geq0,C_9\geq C_4$&
            $1-\Phi_1\left(\sqrt{C_4},0\right)$
            \\ \cline{5-6}
            &&&&
            $C_9<0$&
            $1$
            \\ \cline{3-6}
            &&
            \multirow{7}{*}[0cm]{\rotatebox{90}{$\! C_4\!\geq\! C_7,\! C_4\!<\! C_8$}}&
            $C_5\geq C_7,C_5<C_8$&&
            \tiny{$\!1-\!\Phi_2\!\left(\! \sqrt{C_7},\!0\!\right)\!-\!\frac{2\!\left(\! \sqrt{C_5}\!-\! \sqrt{C_7}\!\right)}{D_y}\!-\!\Phi_3\left(\! \sqrt{C_4},\! \sqrt{C_5},\!\frac{D_y}{2}\!\right)$}
            \\ \cline{4-6}
            &&&
            \multirow{3}{*}[0cm]{$C_5\geq0,C_5<C_7$}&
            $C_9\geq0,C_9<C_7$&
            \tiny{$\!1\!-\!\Phi_2\!\left(\! \sqrt{C_5},\!0\!\right)\!-\!\Phi_1\!\left(\! \sqrt{C_9},\! \sqrt{C_5}\!\right)\!-\!\Phi_3\!\left(\! \sqrt{C_4},\! \sqrt{C_7},\!\frac{D_y}{2}\!\right)$}
            \\ \cline{5-6}
            &&&&
            $C_9\geq0,C_9\geq C_7$&
            \tiny{$\!1\!-\!\Phi_2\!\left(\! \sqrt{C_5},\!0\!\right)\!-\!\Phi_1\!\left(\! \sqrt{C_7},\! \sqrt{C_5}\!\right)\!-\!\Phi_3\!\left(\! \sqrt{C_4},\! \sqrt{C_7},\!\frac{D_y}{2}\!\right)$}
            \\ \cline{5-6}
            &&&&
            $C_9<0$&
            $1-\Phi_2\left(\sqrt{C_5},0\right)-\Phi_3\left(\sqrt{C_4},\sqrt{C_7},\frac{D_y}{2}\right)$
            \\ \cline{4-6}
            &&&
            \multirow{3}{*}[0cm]{$C_5<0$}&
            $C_9\geq0,C_9<C_7$&
            $1-\Phi_1\left(\sqrt{C_9},0\right)-\Phi_3\left(\sqrt{C_4},\sqrt{C_7},\frac{D_y}{2}\right)$
            \\ \cline{5-6}
            &&&&
            $C_9\geq0,C_9\geq C_7$&
            $1-\Phi_1\left(\sqrt{C_7},0\right)-\Phi_3\left(\sqrt{C_4},\sqrt{C_7},\frac{D_y}{2}\right)$
            \\ \cline{5-6}
            &&&&
            $C_9<0$&
            $1-\Phi_3\left(\sqrt{C_4},\sqrt{C_7},\frac{D_y}{2}\right)$
            \\ \cline{3-6}
            &&
            \multirow{8}{*}{\rotatebox{90}{$C_4\geq C_8$}}&
            $C_5\geq C_8$&
            &
            $1-\Phi_2\left(\sqrt{C_7},0\right)-\frac{2\left(\sqrt{C_8}-\sqrt{C_7}\right)}{D_y}$
            \\ \cline{4-6}
            &&&
            $C_5\geq C_7,C_5<C_8$&
            &
            \tiny{$\!1\!-\!\Phi_2\!\left(\! \sqrt{C_7},\!0\!\right)\!-\!\frac{2\!\left(\! \sqrt{C_5}\!-\! \sqrt{C_7}\!\right)}{D_y}\!-\!\Phi_3\!\left(\! \sqrt{C_8},\! \sqrt{C_5},\!\frac{D_y}{2}\!\right)$}
            \\ \cline{4-6}
            &&&
            \multirow{3}{*}[0cm]{$C_5\geq0,C_5<C_7$}&
            $C_9\geq0,C_9<C_7$&
            \tiny{$\!1\!-\!\Phi_2\!\left(\! \sqrt{C_5},\!0\right)\!-\!\Phi_1\!\left(\! \sqrt{C_9},\! \sqrt{C_5}\!\right)\!-\!\Phi_3\!\left(\! \sqrt{C_8},\! \sqrt{C_7},\!\frac{D_y}{2}\!\right)$}
            \\ \cline{5-6}
            &&&&
            $C_9\geq0,C_9\geq C_7$&
            \tiny{$\!1\!-\!\Phi_2\!\left(\! \sqrt{C_5},\!0\right)\!-\!\Phi_1\!\left(\! \sqrt{C_7},\! \sqrt{C_5}\!\right)\!-\!\Phi_3\!\left(\!\sqrt{C_8},\! \sqrt{C_7},\!\frac{D_y}{2}\!\right)$}
            \\ \cline{5-6}
            &&&&
            $C_9<0$&
            $1-\Phi_2\left(\sqrt{C_5},0\right)-\Phi_3\left(\sqrt{C_8},\sqrt{C_7},\frac{D_y}{2}\right)$
            \\ \cline{4-6}
            &&&
            \multirow{3}{*}[0cm]{$C_5<0$}&
            $C_9\geq0,C_9<C_7$&
            $1-\Phi_1\left(\sqrt{C_9},0\right)-\Phi_3\left(\sqrt{C_8},\sqrt{C_7},\frac{D_y}{2}\right)$
            \\ \cline{5-6}
            &&&&
            $C_9\geq0,C_9\geq C_7$&
            $1-\Phi_1\left(\sqrt{C_7},0\right)-\Phi_3\left(\sqrt{C_8},\sqrt{C_7},\frac{D_y}{2}\right)$
            \\ \cline{5-6}
            &&&&
            $C_9<0$&
            $1-\Phi_3\left(\sqrt{C_8},\sqrt{C_7},\frac{D_y}{2}\right)$
            \\ \cline{2-6}
            &
            \multirow{22}{*}[0cm]{\rotatebox{90}{$\sqrt{C_8}\geq\frac{D_y}{2}$}}&
            $C_4<0$&
            &
            &
            $1$
            \\ \cline{3-6}
            &&
            \multirow{6}{*}{\rotatebox{90}{$\! C_4\!\geq\!0\!,\! C_4\!<\! C_7$}}&
            $C_5\geq0,C_5<C_7$&
            $C_9\geq0,C_9<C_4$&
            $1-\Phi_2\left(\sqrt{C_5},0\right)-\Phi_1\left(\sqrt{C_9},\sqrt{C_5}\right)$
            \\ \cline{5-6}
            &&&&
            $C_9\geq0,C_9\geq C_4$&
            $1-\Phi_2\left(\sqrt{C_5},0\right)-\Phi_1\left(\sqrt{C_4},\sqrt{C_5}\right)$
            \\ \cline{5-6}
            &&&&
            $C_9<0$&
            $1-\Phi_2\left(\sqrt{C_5},0\right)$
            \\ \cline{4-6}
            &&&
            \multirow{3}{*}[0cm]{$C_5<0$}&
            $C_9\geq0,C_9<C_4$&
            $1-\Phi_1\left(\sqrt{C_9},0\right)$
            \\ \cline{5-6}
            &&&&			
            $C_9\geq0,C_9\geq C_4$&
            $1-\Phi_1\left(\sqrt{C_4},0\right)$
            \\ \cline{5-6}
            &&&&
            $C_9<0$&
            $1$
            \\ \cline{3-6}
            &&
            \multirow{7}{*}[0cm]{\rotatebox{90}{$\! C_4\!\geq\! C_7,\! \sqrt{C_4}\!<\! \frac{D_y}{2}$}}&
            $C_5\geq C_7,\sqrt{C_5}<\frac{D_y}{2}$&&
            \tiny{$\!1-\!\Phi_2\!\left(\! \sqrt{C_7},\!0\!\right)\!-\!\frac{2\!\left(\! \sqrt{C_5}\!-\! \sqrt{C_7}\!\right)}{D_y}\!-\!\Phi_3\left(\! \sqrt{C_4},\! \sqrt{C_5},\!\frac{D_y}{2}\!\right)$}
            \\ \cline{4-6}
            &&&
            \multirow{3}{*}[0cm]{$C_5\geq0,C_5<C_7$}&
            $C_9\geq0,C_9<C_7$&
            \tiny{$\!1\!-\!\Phi_2\!\left(\! \sqrt{C_5},\!0\!\right)\!-\!\Phi_1\!\left(\! \sqrt{C_9},\! \sqrt{C_5}\!\right)\!-\!\Phi_3\!\left(\! \sqrt{C_4},\! \sqrt{C_7},\!\frac{D_y}{2}\!\right)$}
            \\ \cline{5-6}
            &&&&
            $C_9\geq0,C_9\geq C_7$&
            \tiny{$\!1\!-\!\Phi_2\!\left(\! \sqrt{C_5},\!0\!\right)\!-\!\Phi_1\!\left(\! \sqrt{C_7},\! \sqrt{C_5}\!\right)\!-\!\Phi_3\!\left(\! \sqrt{C_4},\! \sqrt{C_7},\!\frac{D_y}{2}\!\right)$}
            \\ \cline{5-6}
            &&&&
            $C_9<0$&
            $1-\Phi_2\left(\sqrt{C_5},0\right)-\Phi_3\left(\sqrt{C_4},\sqrt{C_7},\frac{D_y}{2}\right)$
            \\ \cline{4-6}
            &&&
            \multirow{3}{*}[0cm]{$C_5<0$}&
            $C_9\geq0,C_9<C_7$&
            $1-\Phi_1\left(\sqrt{C_9},0\right)-\Phi_3\left(\sqrt{C_4},\sqrt{C_7},\frac{D_y}{2}\right)$
            \\ \cline{5-6}
            &&&&
            $C_9\geq0,C_9\geq C_7$&
            $1-\Phi_1\left(\sqrt{C_7},0\right)-\Phi_3\left(\sqrt{C_4},\sqrt{C_7},\frac{D_y}{2}\right)$
            \\ \cline{5-6}
            &&&&
            $C_9<0$&
            $1-\Phi_3\left(\sqrt{C_4},\sqrt{C_7},\frac{D_y}{2}\right)$
            \\ \cline{3-6}
            &&
            \multirow{8}{*}[0cm]{\rotatebox{90}{$\sqrt{C_4}\geq\frac{D_y}{2}$}}&
            $\sqrt{C_5}\geq\frac{D_y}{2}$&
            &
            $1-\Phi_2\left(\sqrt{C_7},0\right)$
            \\ \cline{4-6}
            &&&
            $C_5\geq C_7,\sqrt{C_5}<\frac{D_y}{2}$&
            &
            \tiny{$\!1\!-\!\Phi_2\!\left(\! \sqrt{C_7},\!0\!\right)\!-\!\frac{2\!\left(\! \sqrt{C_5}\!-\! \sqrt{C_7}\!\right)}{D_y}\!-\!\Phi_3\!\left(\!\frac{D_y}{2},\! \sqrt{C_5},\!\frac{D_y}{2}\!\right)$}
            \\ \cline{4-6}
            &&&
            \multirow{3}{*}[0cm]{$C_5\geq0,C_5<C_7$}&
            $C_9^2\geq0,C_9<C_7$&
            \tiny{$\!1\!-\!\Phi_2\!\left(\! \sqrt{C_5},\!0\!\right)\!-\!\Phi_1\!\left(\! \sqrt{C_9},\! \sqrt{C_5}\!\right)\!-\!\Phi_3\!\left(\!\frac{D_y}{2},\! \sqrt{C_7},\!\frac{D_y}{2}\!\right)$}
            \\ \cline{5-6}
            &&&&
            $C_9\geq0,C_9\geq C_7$&
            \tiny{$\!1\!-\!\Phi_2\!\left(\! \sqrt{C_5},\!0\!\right)\!-\!\Phi_1\!\left(\! \sqrt{C_7},\! \sqrt{C_5}\!\right)\!-\!\Phi_3\!\left(\!\frac{D_y}{2},\! \sqrt{C_7},\!\frac{D_y}{2}\!\right)$}
            \\ \cline{5-6}
            &&&&
            $C_9<0$&
            $1-\Phi_2\left(\sqrt{C_5},0\right)-\Phi_3\left(\frac{D_y}{2},\sqrt{C_7},\frac{D_y}{2}\right)$
            \\ \cline{4-6}
            &&&
            \multirow{3}{*}[0cm]{$C_5<0$}&
            $C_9\geq0,C_9<C_7$&
            $1-\Phi_1\left(\sqrt{C_9},0\right)-\Phi_3\left(\frac{D_y}{2},0,\frac{D_y}{2}\right)$
            \\ \cline{5-6}
            &&&&
            $C_9\geq0,C_9\geq C_7$&
            $1-\Phi_1\left(\sqrt{C_7},0\right)-\Phi_3\left(\frac{D_y}{2},0,\frac{D_y}{2}\right)$
            \\ \cline{5-6}
            &&&&
            $C_9<0$&
            $1-\Phi_3\left(\frac{D_y}{2},0,\frac{D_y}{2}\right)$
            \\ \hline \hline
        \end{tabular}
    \end{table*}
    \begin{table*}
        \caption{Expressions for OP $x_{2a}$ $C_6<0$ v3}
        \label{OP2av3}
        \centering
        \begin{tabular}{c|c|c|c|c|c}
            \hline \hline
            \multicolumn{5}{c|}{Conditions}
            & Expression
            \\ \hline \hline
            \multirow{29}{*}[0cm]{\rotatebox{90}{$\sqrt{C_7} \geq \frac{D_y}{2}$}} &
            $C_8<0$&
            &
            &
            &
            $1$
            \\ \cline{2-6}
            &
            \multirow{14}{*}[0cm]{\rotatebox{90}{$C_8\geq0,\sqrt{C_8}<\frac{D_y}{2}$}}&
            $C_4<0$&
            &
            &
            1
            \\ \cline{3-6}
            &&
            \multirow{6}{*}[0cm]{\rotatebox{90}{$\! C_4\!\geq\!0,\! C_4\!<\! C_8$}}&
            \multirow{3}{*}[0cm]{$C_5\geq0,C_5<C_8$}&
            $C_9\geq0,C_9<C_4$&
            $1-\Phi_2\left(\sqrt{C_5},0\right)-\Phi_1\left(\sqrt{C_9},\sqrt{C_5}\right)$
            \\ \cline{5-6}
            &&&&
            $C_9\geq0,C_9\geq C_4$&
            $1-\Phi_2\left(\sqrt{C_5},0\right)-\Phi_1\left(\sqrt{C_4},\sqrt{C_5}\right)$
            \\ \cline{5-6}
            &&&&
            $C_9<0$&
            $1-\Phi_2\left(\sqrt{C_5},0\right)$
            \\ \cline{4-6}
            &&&
            \multirow{3}{*}[0cm]{$C_5<0$}&
            $C_9\geq0,C_9<C_4$&
            $1-\Phi_1\left(\sqrt{C_9},0\right)$
            \\ \cline{5-6}
            &&&&
            $C_9\geq0,C_9\geq C_4$&
            $1-\Phi_1\left(\sqrt{C_4},0\right)$
            \\ \cline{5-6}
            &&&&
            $C_9<0$&
            $1$
            \\ \cline{3-6}
            &&
            \multirow{7}{*}[0cm]{\rotatebox{90}{$C_4\geq C_8$}}&
            $C_5\geq C_8$&
            &
            $1-\Phi_2\left(\sqrt{C_8},0\right)$
            \\ \cline{4-6}
            &&&
            \multirow{3}{*}[0cm]{$C_5\geq0,C_5<C_8$}&
            $C_9\geq0,C_9<C_8$&
            $1-\Phi_2\left(\sqrt{C_5},0\right)-\Phi_1\left(\sqrt{C_9},\sqrt{C_5}\right)$
            \\ \cline{5-6}
            &&&&
            $C_9\geq0,C_9\geq C_8$&
            $1-\Phi_2\left(\sqrt{C_5},0\right)-\Phi_1\left(\sqrt{C_8},\sqrt{C_5}\right)$
            \\ \cline{5-6}
            &&&&
            $C_9<0$&
            $1-\Phi_2\left(\sqrt{C_5},0\right)$
            \\ \cline{4-6}
            &&&
            \multirow{3}{*}[0cm]{$C_5<0$}&
            $C_9\geq0,C_9<C_8$&
            $1-\Phi_1\left(\sqrt{C_9},0\right)$
            \\ \cline{5-6}
            &&&&
            $C_9\geq0,C_9\geq C_8$&
            $1-\Phi_1\left(\sqrt{C_8},0\right)$
            \\ \cline{5-6}
            &&&&
            $C_9<0$&
            $1$
            \\ \cline{2-6}
            &
            \multirow{14}{*}[0cm]{\rotatebox{90}{$\sqrt{C_8}\geq\frac{D_y}{2}$}}&
            $C_4<0$&
            &&
            $1$
            \\ \cline{3-6}
            &&
            \multirow{6}{*}[0cm]{\rotatebox{90}{$\! C_4\!\geq\!0,\! \sqrt{C_4}\!<\!\frac{D_y}{2}$}}&
            \multirow{3}{*}[0cm]{$C_5\geq0,\sqrt{C_5}<\frac{D_y}{2}$}&
            $C_9\geq0,C_9<C_4$&
            $1-\Phi_2\left(\sqrt{C_5},0\right)-\Phi_1\left(\sqrt{C_9},\sqrt{C_5}\right)$
            \\ \cline{5-6}
            &&&&
            $C_9\geq0,C_9\geq C_4$&
            $1-\Phi_2\left(\sqrt{C_5},0\right)-\Phi_1\left(\sqrt{C_4},\sqrt{C_5}\right)$
            \\ \cline{5-6}
            &&&&
            $C_9<0$&
            $1-\Phi_2\left(\sqrt{C_5},0\right)$
            \\ \cline{4-6}
            &&&
            \multirow{3}{*}[0cm]{$C_5<0$}&
            $C_9\geq0,C_9<C_4$&
            $1-\Phi_1\left(\sqrt{C_9},0\right)$
            \\ \cline{5-6}
            &&&&
            $C_9\geq0,C_9\geq C_4$&
            $1-\Phi_1\left(\sqrt{C_4},0\right)$
            \\ \cline{5-6}
            &&&&
            $C_9<0$&
            $1$
            \\ \cline{3-6}
            &&
            \multirow{7}{*}[0cm]{\rotatebox{90}{$\sqrt{C_4}\geq\frac{D_y}{2}$}}&
            $\sqrt{C_5}\geq\frac{D_y}{2}$&
            &
            $1-\Phi_2\left(\frac{D_y}{2},0\right)$
            \\ \cline{4-6}
            &&&
            \multirow{3}{*}[0cm]{$C_5\geq0,\sqrt{C_5}<\frac{D_y}{2}$}&
            $C_9\geq0,\sqrt{C_9}<\frac{D_y}{2}$&
            $1-\Phi_2\left(\sqrt{C_5},0\right)-\Phi_1\left(\sqrt{C_9},\sqrt{C_5}\right)$
            \\ \cline{5-6}
            &&&&
            $C_9\geq0,\sqrt{C_9}\geq\frac{D_y}{2}$&
            $1-\Phi_2\left(\sqrt{C_5},0\right)-\Phi_1\left(\frac{D_y}{2},\sqrt{C_5}\right)$
            \\ \cline{5-6}
            &&&&
            $C_9<0$&
            $1-\Phi_2\left(\sqrt{C_5},0\right)$
            \\ \cline{4-6}
            &&&
            \multirow{3}{*}[0cm]{$C_5<0$}&
            $C_9\geq0,\sqrt{C_9}<\frac{D_y}{2}$&
            $1-\Phi_1\left(\sqrt{C_9},0\right)$
            \\ \cline{5-6}
            &&&&
            $C_9\geq0,\sqrt{C_9}\geq\frac{D_y}{2}$&
            $1-\Phi_1\left(\frac{D_y}{2},0\right)$
            \\ \cline{5-6}
            &&&&
            $C_9<0$&
            $1$
            \\ \hline \hline
        \end{tabular}
    \end{table*}
    \begin{table*}
        \caption{Expressions for OP $x_{2a}$ $C_6\geq0,\sqrt{C_6}<\frac{D_y}{2}$}
        \label{OP2av4}
        \centering
        \begin{tabular}{c|c|c|c|c|c}
            \hline \hline
            \multicolumn{5}{c|}{Conditions}
            & Expression
            \\ \hline \hline
            \multirow{29}{*}[0cm]{\rotatebox{90}{$C_7 < 0$}} &
            $C_8<C_6$&
            &
            &
            &
            $1$
            \\ \cline{2-6}
            &
            \multirow{14}{*}[0cm]{\rotatebox{90}{$C_8 \geq 0,\sqrt{C_8}<\frac{D_y}{2}$}} &
            $C_4<C_6$&
            &
            &
            1
            \\ \cline{3-6}
            &&
            \multirow{6}{*}[0cm]{\rotatebox{90}{$\! C_4\!\geq\! C_6\!,\! C_4\!\!<\!\! C_8\!$}}&
            \multirow{3}{*}[0cm]{$C_5\geq C_6,C_5<C_8$}&
            $C_9\geq0,C_9<C_4$&
            $1-\Phi_2\left(\sqrt{C_5},\sqrt{C_6}\right)-\Phi_1\left(\sqrt{C_9},\sqrt{C_5}\right)$
            \\ \cline{5-6}
            &&&&
            $C_9\geq0,C_9\geq C_4$&
            $1-\Phi_2\left(\sqrt{C_5},\sqrt{C_6}\right)-\Phi_1\left(\sqrt{C_4},\sqrt{C_5}\right)$
            \\ \cline{5-6}
            &&&&
            $C_9<0$&
            $1-\Phi_2\left(\sqrt{C_5},\sqrt{C_6}\right)$
            \\ \cline{4-6}
            &&&
            \multirow{3}{*}[0cm]{$C_5<C_6$}&
            $C_9\geq0,C_9<C_4$&
            $1-\Phi_1\left(\sqrt{C_9},\sqrt{C_6}\right)$
            \\ \cline{5-6}
            &&&&
            $C_9\geq0,C_9\geq C_4$&
            $1-\Phi_1\left(\sqrt{C_4},\sqrt{C_6}\right)$
            \\ \cline{5-6}
            &&&&
            $C_9<0$&
            $1$
            \\ \cline{3-6}
            &&
            \multirow{7}{*}[0cm]{\rotatebox{90}{$C_4\geq C_8$}}&
            $C_5\geq C_8$&
            &
            $1-\Phi_2\left(\sqrt{C_8},\sqrt{C_6}\right)$
            \\ \cline{4-6}
            &&&
            \multirow{3}{*}[0cm]{$C_5\geq C_6,C_5<C_8$}&
            $C_9\geq0,C_9<C_8$&
            $1-\Phi_2\left(\sqrt{C_5},\sqrt{C_6}\right)-\Phi_1\left(\sqrt{C_9},\sqrt{C_5}\right)$
            \\ \cline{5-6}
            &&&&
            $C_9\geq0,C_9\geq C_8$&
            $1-\Phi_2\left(\sqrt{C_5},\sqrt{C_6}\right)-\Phi_1\left(\sqrt{C_8},\sqrt{C_5}\right)$
            \\ \cline{5-6}
            &&&&
            $C_9<0$&
            $1-\Phi_2\left(\sqrt{C_5},\sqrt{C_6}\right)$
            \\ \cline{4-6}
            &&&
            \multirow{3}{*}[0cm]{$C_5<C_6$}&
            $C_9\geq0,C_9<C_8$&
            $1-\Phi_1\left(\sqrt{C_9},\sqrt{C_6}\right)$
            \\ \cline{5-6}
            &&&&
            $C_9\geq0,C_9\geq C_8$&
            $1-\Phi_1\left(\sqrt{C_8},\sqrt{C_6}\right)$
            \\ \cline{5-6}
            &&&&
            $C_9<0$&
            $1$
            \\ \cline{2-6}
            &
            \multirow{14}{*}[0cm]{\rotatebox{90}{$\sqrt{C_8}\geq\frac{D_y}{2}$}}&
            $C_4<C_6$&
            &&
            $1$
            \\ \cline{3-6}
            &&
            \multirow{6}{*}[0cm]{\rotatebox{90}{$\! C_4\! \geq\! C_6\!,\! \sqrt{C_4}\!\!<\! \!\frac{D_y}{2}\!$}}&
            \multirow{3}{*}[0cm]{$C_5\geq C_6,\sqrt{C_5}<\frac{D_y}{2}$}&
            $C_9\geq0,C_9<C_4$&
            $1-\Phi_2\left(\sqrt{C_5},\sqrt{C_6}\right)-\Phi_1\left(\sqrt{C_9},\sqrt{C_5}\right)$
            \\ \cline{5-6}
            &&&&
            $C_9\geq0,C_9\geq C_4$&
            $1-\Phi_2\left(\sqrt{C_5},\sqrt{C_6}\right)-\Phi_1\left(\sqrt{C_4},\sqrt{C_5}\right)$
            \\ \cline{5-6}
            &&&&
            $C_9<0$&
            $1-\Phi_2\left(\sqrt{C_5},\sqrt{C_6}\right)$
            \\ \cline{4-6}
            &&&
            \multirow{3}{*}[0cm]{$C_5<C_6$}&
            $C_9\geq0,C_9<C_4$&
            $1-\Phi_1\left(\sqrt{C_9},\sqrt{C_6}\right)$
            \\ \cline{5-6}
            &&&&
            $C_9\geq0,C_9\geq C_4$&
            $1-\Phi_1\left(\sqrt{C_4},\sqrt{C_6}\right)$
            \\ \cline{5-6}
            &&&&
            $C_9<0$&
            $1$
            \\ \cline{3-6}
            &&
            \multirow{7}{*}[0cm]{\rotatebox{90}{$\sqrt{C_4}\geq\frac{D_y}{2}$}}&
            $\sqrt{C_5}\geq\frac{D_y}{2}$&
            &
            $1-\Phi_2\left(\frac{D_y}{2},\sqrt{C_6}\right)$
            \\ \cline{4-6}
            &&&
            \multirow{3}{*}[0cm]{$C_5\geq C_6, \sqrt{C_5}<\frac{D_y}{2}$}&
            $C_9\geq0,\sqrt{C_9}<\frac{D_y}{2}$&
            $1-\Phi_2\left(\sqrt{C_5},\sqrt{C_6}\right)-\Phi_1\left(\sqrt{C_9},\sqrt{C_5}\right)$
            \\ \cline{5-6}
            &&&&
            $C_9\geq0,\sqrt{C_9}\geq \frac{D_y}{2}$&
            $1-\Phi_2\left(\sqrt{C_5},\sqrt{C_6}\right)-\Phi_1\left(\frac{D_y}{2},\sqrt{C_5}\right)$
            \\ \cline{5-6}
            &&&&
            $C_9<0$&
            $1-\Phi_2\left(\sqrt{C_5},\sqrt{C_6}\right)$
            \\ \cline{4-6}
            &&&
            \multirow{3}{*}[0cm]{$C_5<C_6$}&
            $C_9\geq0,\sqrt{C_9}<\frac{D_y}{2}$&
            $1-\Phi_1\left(\sqrt{C_9},\sqrt{C_6}\right)$
            \\ \cline{5-6}
            &&&&
            $C_9\geq0,\sqrt{C_9}\geq\frac{D_y}{2}$&
            $1-\Phi_1\left(\frac{D_y}{2},\sqrt{C_6}\right)$
            \\ \cline{5-6}
            &&&&
            $C_9<0$&
            $1$
            \\ \hline \hline
        \end{tabular}
    \end{table*}
    \begin{table*}
        \caption{Expressions for OP $x_{2a}$ $C_6\geq0,\sqrt{C_6}<\frac{D_y}{2}$}
        \label{OP2av5}
        \centering
        \begin{tabular}{c|c|c|c|c|c}
            \hline \hline
            \multicolumn{5}{c|}{Conditions}
            & Expression
            \\ \hline \hline
            \multirow{59}{*}[0cm]{\rotatebox{90}{$C_7 \geq 0,\sqrt{C_7}<\frac{D_y}{2}$}} &
            $C_8<C_6$&
            &
            &
            &
            $1$
            \\ \cline{2-6}
            &
            \multirow{14}{*}[0cm]{\rotatebox{90}{$C_8 \geq C_6,C_8<C_7$}} &
            $C_4<C_6$&
            &
            &
            $1$
            \\ \cline{3-6}
            &&
            \multirow{6}{*}[0cm]{\rotatebox{90}{$\! C_4\!\!\geq\!\! C_6,\! C_4\!<\! C_8$}}&
            \multirow{3}{*}[0cm]{$C_5\geq C_6,C_5<C_8$}&
            $C_9\geq0,C_9<C_4$&
            $1-\Phi_2\left(\sqrt{C_5},\sqrt{C_6}\right)-\Phi_1\left(\sqrt{C_9},\sqrt{C_5}\right)$
            \\ \cline{5-6}
            &&&&
            $C_9\geq0,C_9\geq C_4$&
            $1-\Phi_2\left(\sqrt{C_5},\sqrt{C_6}\right)-\Phi_1\left(\sqrt{C_4},\sqrt{C_5}\right)$
            \\ \cline{5-6}
            &&&&
            $C_9<0$&
            $1-\Phi_2\left(\sqrt{C_5},\sqrt{C_6}\right)$
            \\ \cline{4-6}
            &&&
            \multirow{3}{*}[0cm]{$C_5<C_6$}&
            $C_9\geq0,C_9<C_4$&
            $1-\Phi_1\left(\sqrt{C_9},\sqrt{C_6}\right)$
            \\ \cline{5-6}
            &&&&
            $C_9\geq0,C_9\geq C_4$&
            $1-\Phi_1\left(\sqrt{C_3},\sqrt{C_6}\right)$
            \\ \cline{5-6}
            &&&&
            $C_9<0$&
            $1$
            \\ \cline{3-6}
            &&
            \multirow{7}{*}[0cm]{$C_4\geq C_8$}&
            $C_5\geq C_8$&
            &
            $1-\Phi_2\left(\sqrt{C_8},\sqrt{C_6}\right)$
            \\ \cline{4-6}
            &&&
            \multirow{3}{*}[0cm]{$C_5\geq C_6, C_5<C_8$}&
            $C_9\geq0,C_9<C_8$&
            $1-\Phi_2\left(\sqrt{C_5},\sqrt{C_6}\right)-\Phi_1\left(\sqrt{C_9},\sqrt{C_5}\right)$
            \\ \cline{5-6}
            &&&&
            $C_9\geq0,C_9\geq C_8$&
            $1-\Phi_2\left(\sqrt{C_5},\sqrt{C_6}\right)-\Phi_1\left(\sqrt{C_8},\sqrt{C_5}\right)$
            \\ \cline{5-6}
            &&&&
            $C_9<0$&
            $1-\Phi_2\left(\sqrt{C_5},\sqrt{C_6}\right)$
            \\ \cline{4-6}
            &&&
            \multirow{3}{*}[0cm]{$C_5<C_6$}&
            $C_9\geq0,C_9<C_8$&
            $1-\Phi_1\left(\sqrt{C_9},\sqrt{C_6}\right)$
            \\ \cline{5-6}
            &&&&
            $C_9\geq0,C_9\geq C_8$&
            $1-\Phi_1\left(\sqrt{C_8},\sqrt{C_6}\right)$
            \\ \cline{5-6}
            &&&&
            $C_9<0$&
            $1$
            \\ \cline{2-6}
            &
            \multirow{22}{*}[0cm]{\rotatebox{90}{$C_8\geq C_7, \sqrt{C_8}<\frac{D_y}{2}$}}&
            $C_4<C_6$&
            &&
            $1$
            \\ \cline{3-6}
            &&
            \multirow{6}{*}[0cm]{\rotatebox{90}{$\! C_4\!\!\geq\!\! C_6,\! C_4\!\!<\!\! C_7$}}&
            \multirow{3}{*}[0cm]{$C_5\geq C_6, C_5<C_7$}&
            $C_9\geq0,C_9<C_4$&
            $1-\Phi_2\left(\sqrt{C_5},\sqrt{C_6}\right)-\Phi_1\left(\sqrt{C_9},\sqrt{C_5}\right)$
            \\ \cline{5-6}
            &&&&
            $C_9\geq0,C_9\geq C_4$&
            $1-\Phi_2\left(\sqrt{C_5},\sqrt{C_6}\right)-\Phi_1\left(\sqrt{C_4},\sqrt{C_5}\right)$
            \\ \cline{5-6}
            &&&&
            $C_9<0$&
            $1-\Phi_2\left(\sqrt{C_5},\sqrt{C_6}\right)$
            \\ \cline{4-6}
            &&&
            \multirow{3}{*}[0cm]{$C_5<C_6$}&
            $C_9\geq 0,C_9<C_4$&
            $1-\Phi_1\left(\sqrt{C_9},\sqrt{C_6}\right)$
            \\ \cline{5-6}
            &&&&
            $C_9\geq 0,C_9\geq C_4$&
            $1-\Phi_1\left(\sqrt{C_4},\sqrt{C_6}\right)$
            \\ \cline{5-6}
            &&&&
            $C_9< 0$&
            $1$
            \\ \cline{3-6}
            &&
            \multirow{7}{*}[0cm]{\rotatebox{90}{$C_4\geq C_7,C_4<C_8$}}&
            $C_5\geq C_7,C_5<C_8$&&
            \tiny{$\!1\!-\!\Phi_2\!\left(\! \sqrt{C_7},\! \sqrt{C_6}\!\right)\!-\!\frac{2\left(\! \sqrt{C_5}\!-\! \sqrt{C_7}\!\right)}{D_y}\!-\!\Phi_3\!\left(\! \sqrt{C_4},\! \sqrt{C_5},\! \frac{D_y}{2}\!\right)$}
            \\ \cline{4-6}
            &&&
            \multirow{3}{*}[0cm]{$C_5\geq C_6,C_5<C_7$}&
            $C_9\geq0,C_9<C_7$&
            \tiny{$\!1\!-\!\Phi_2\!\left(\! \sqrt{C_5},\! \sqrt{C_6}\!\right)\!-\!\Phi_1\left(\! \sqrt{C_9},\! \sqrt{C_5}\!\right)\!-\!\Phi_3\!\left(\! \sqrt{C_4},\! \sqrt{C_7},\! \frac{D_y}{2}\!\right)$}
            \\ \cline{5-6}
            &&&&
            $C_9\geq0,C_9\geq C_7$&
            \tiny{$\!1\!-\!\Phi_2\!\left(\! \sqrt{C_5},\! \sqrt{C_6}\!\right)\!-\!\Phi_1\left(\! \sqrt{C_7},\! \sqrt{C_5}\!\right)\!-\!\Phi_3\!\left(\! \sqrt{C_4},\! \sqrt{C_7},\! \frac{D_y}{2}\!\right)$}
            \\ \cline{5-6}
            &&&&
            $C_9<0$&
            $1-\Phi_2\left(\sqrt{C_5},\sqrt{C_6}\right)-\Phi_3\left(\sqrt{C_4},\sqrt{C_7},\frac{D_y}{2}\right)$
            \\ \cline{4-6}
            &&&
            \multirow{3}{*}[0cm]{$C_5<C_6$}&
            $C_9\geq0,C_9<C_7$&
            $1-\Phi_1\left(\sqrt{C_9},\sqrt{C_6}\right)-\Phi_3\left(\sqrt{C_4},\sqrt{C_7},\frac{D_y}{2}\right)$
            \\ \cline{5-6}
            &&&&
            $C_9\geq0,C_9\geq C_7$&
            $1-\Phi_1\left(\sqrt{C_7},\sqrt{C_6}\right)-\Phi_3\left(\sqrt{C_4},\sqrt{C_7},\frac{D_y}{2}\right)$
            \\ \cline{5-6}
            &&&&
            $C_9<0$&
            $1-\Phi_3\left(\sqrt{C_4},\sqrt{C_7},\frac{D_y}{2}\right)$
            \\ \cline{3-6}
            &&
            \multirow{8}{*}[0cm]{\rotatebox{90}{$C_4\geq C_8$}}&
            $C_5\geq C_8$&&
            $1-\Phi_2\left(\sqrt{C_7},\sqrt{C_6}\right)-\frac{2\left(\sqrt{C_8}-\sqrt{C_7}\right)}{D_y}$
            \\ \cline{4-6}
            &&&
            $C_5\geq C_7,C_5<C_8$&&
            $1-\Phi_2\left(\sqrt{C_7},\sqrt{C_6}\right)-\frac{2\left(\sqrt{C_5}-\sqrt{C_7}\right)}{D_y}$
            \\ \cline{4-6}
            &&&
            \multirow{3}{*}[0cm]{$C_5\geq C_6,C_5<C_7$}&
            $C_9\geq0, C_9<C_7$&
            \tiny{$\!1\!-\!\Phi_2\!\left(\! \sqrt{C_5},\! \sqrt{C_6}\!\right)\!-\!\Phi_1\!\left(\! \sqrt{C_9},\! \sqrt{C_5}\!\right)\!-\!\Phi_3\!\left(\! \sqrt{C_8},\! \sqrt{C_7},\!\frac{D_y}{2}\!\right)$}
            \\ \cline{5-6}
            &&&&
            $C_9\geq0, C_9\geq C_7$&
            \tiny{$\!1\!-\!\Phi_2\!\left(\! \sqrt{C_5},\! \sqrt{C_6}\!\right)\!-\!\Phi_1\!\left(\! \sqrt{C_7},\! \sqrt{C_5}\!\right)\!-\!\Phi_3\!\left(\! \sqrt{C_8},\! \sqrt{C_7},\!\frac{D_y}{2}\!\right)$}
            \\ \cline{5-6}
            &&&&
            $C_9<0$&
            $1-\Phi_2\left(\sqrt{C_5},\sqrt{C_6}\right)-\Phi_3\left(\sqrt{C_8},\sqrt{C_7},\frac{D_y}{2}\right)$
            \\ \cline{4-6}
            &&&
            \multirow{3}{*}[0cm]{$C_5<C_6$}&
            $C_9\geq0,C_9<C_7$&
            $1-\Phi_1\left(\sqrt{C_9},\sqrt{C_6}\right)-\Phi_3\left(\sqrt{C_8},\sqrt{C_7},\frac{D_y}{2}\right)$
            \\ \cline{5-6}
            &&&&
            $C_9\geq0,C_9\geq C_7$&
            $1-\Phi_1\left(\sqrt{C_7},\sqrt{C_6}\right)-\Phi_3\left(\sqrt{C_8},\sqrt{C_7},\frac{D_y}{2}\right)$
            \\ \cline{5-6}
            &&&&
            $C_9<0$&
            $1-\Phi_3\left(\sqrt{C_8},\sqrt{C_7},\frac{D_y}{2}\right)$
            \\ \cline{2-6}
            &
            \multirow{22}{*}[0cm]{\rotatebox{90}{$\sqrt{C_8}\geq\frac{D_y}{2}$}}&
            $C_4<C_6$&&&
            $1$
            \\ \cline{3-6}
            &&
            \multirow{6}{*}[0cm]{\rotatebox{90}{$\! C_4\!\!\geq\!\! C_6,\! C_4\!<\! C_7$}}&
            \multirow{3}{*}[0cm]{$C_5\geq C_6,C_5<C_7$}&
            $C_9\geq0, C_9<C_4$&
            $1-\Phi_2\left(\sqrt{C_5},\sqrt{C_6}\right)-\Phi_1\left(\sqrt{C_9},\sqrt{C_5}\right)$
            \\ \cline{5-6}
            &&&&
            $C_9\geq0, C_9\geq C_4$&
            $1-\Phi_2\left(\sqrt{C_5},\sqrt{C_6}\right)-\Phi_1\left(\sqrt{C_4},\sqrt{C_5}\right)$
            \\ \cline{5-6}
            &&&&
            $C_9<0$&
            $1-\Phi_2\left(\sqrt{C_5},\sqrt{C_6}\right)$
            \\ \cline{4-6}
            &&&
            \multirow{3}{*}[0cm]{$C_5<C_6$}&
            $C_9\geq0,C_9<C_4$&
            $1-\Phi_1\left(\sqrt{C_9},\sqrt{C_6}\right)$
            \\ \cline{5-6}
            &&&&
            $C_9\geq0,C_9\geq C_4$&
            $1-\Phi_1\left(\sqrt{C_3},\sqrt{C_6}\right)$
            \\ \cline{5-6}
            &&&&
            $C_9<0$&
            $1$
            \\ \cline{3-6}
            &&
            \multirow{7}{*}[0cm]{\rotatebox{90}{$C_4\geq C_7,\sqrt{C_4}<\frac{D_y}{2}$}}&
            $C_5\geq C_7,\sqrt{C_5}<\frac{D_y}{2}$&&
            \tiny{$\!1\!-\!\Phi_2\!\left(\! \sqrt{C_7},\! \sqrt{C_6}\!\right)\!-\!\frac{2\!\left(\! \sqrt{C_5}\!-\! \sqrt{C_7}\!\right)}{D_y}\!-\!\Phi_3\!\left(\! \sqrt{C_4},\! \sqrt{C_5},\!\frac{D_y}{2}\!\right)$}
            \\ \cline{4-6}
            &&&
            \multirow{3}{*}[0cm]{$C_5\geq C_6, C_5<C_7$}&
            $C_9\geq 0,C_9<C_7$&
            \tiny{$\!1\!-\!\Phi_2\!\left(\! \sqrt{C_5},\! \sqrt{C_6}\!\right)\!-\!\Phi_1\!\left(\! \sqrt{C_9},\! \sqrt{C_5}\!\right)\!-\!\Phi_3\!\left(\! \sqrt{C_4},\! \sqrt{C_7},\!\frac{D_y}{2}\!\right)$}
            \\ \cline{5-6}
            &&&&
            $C_9\geq 0,C_9\geq C_7$&
            \tiny{$\!1\!-\!\Phi_2\!\left(\! \sqrt{C_5},\! \sqrt{C_6}\!\right)\!-\!\Phi_1\!\left(\! \sqrt{C_7},\! \sqrt{C_5}\!\right)\!-\!\Phi_3\!\left(\! \sqrt{C_4},\! \sqrt{C_7},\!\frac{D_y}{2}\!\right)$}
            \\ \cline{5-6}
            &&&&
            $C_9<0$&
            $1-\Phi_2\left(\sqrt{C_5},\sqrt{C_6}\right)-\Phi_3\left(\sqrt{C_4},\sqrt{C_7},\frac{D_y}{2}\right)$
            \\ \cline{4-6}
            &&&
            \multirow{3}{*}[0cm]{$C_5<C_6$}&
            $C_9\geq 0,C_9<C_7$&
            $1-\Phi_1\left(\sqrt{C_9},\sqrt{C_6}\right)-\Phi_3\left(\sqrt{C_4},\sqrt{C_7},\frac{D_y}{2}\right)$
            \\ \cline{5-6}
            &&&&
            $C_9\geq 0,C_9\geq C_7$&
            $1-\Phi_1\left(\sqrt{C_7},\sqrt{C_6}\right)-\Phi_3\left(\sqrt{C_4},\sqrt{C_7},\frac{D_y}{2}\right)$
            \\ \cline{5-6}
            &&&&
            $C_9<0$&
            $1-\Phi_3\left(\sqrt{C_4},\sqrt{C_7},\frac{D_y}{2}\right)$
            \\ \cline{3-6}
            &&
            \multirow{8}{*}[0cm]{$\sqrt{C_4}\geq\frac{D_y}{2}$}&
            $\sqrt{C_5}\geq\frac{D_y}{2}$&&
            $1-\Phi_2\left(\sqrt{C_7},\sqrt{C_6}\right)-\frac{2}{D_y}\left(\frac{D_y}{2}-\sqrt{C_7}\right)$
            \\ \cline{4-6}
            &&&
            $C_5\geq C_7,\sqrt{C_5}<\frac{D_y}{2}$&&
            \tiny{$\!1\!-\!\Phi_2\!\left(\! \sqrt{C_7},\! \sqrt{C_6}\!\right)\!-\!\frac{2\!\left(\! \sqrt{C_5}\!-\! \sqrt{C_7}\!\right)}{D_y}\!-\!\Phi_3\!\left(\!\frac{D_y}{2},\! \sqrt{C_5},\!\frac{D_y}{2}\right)$}
            \\ \cline{4-6}
            &&&
            $C_5\geq C_6, C_5<C_7$&
            $C_9\geq0,C_9<C_7$&
            \tiny{$\!1\!-\!\Phi_2\!\left(\! \sqrt{C_5},\! \sqrt{C_6}\!\right)\!-\!\Phi_1\!\left(\! \sqrt{C_9},\! \sqrt{C_5}\!\right)\!-\!\Phi_3\!\left(\!\frac{D_y}{2},\! \sqrt{C_5},\!\frac{D_y}{2}\right)$}
            \\ \cline{5-6}
            &&&&
            $C_9\geq0,C_9\geq C_7$&
            \tiny{$\!1\!-\!\Phi_2\!\left(\! \sqrt{C_5},\! \sqrt{C_6}\!\right)\!-\!\Phi_1\!\left(\! \sqrt{C_7},\! \sqrt{C_5}\!\right)\!-\!\Phi_3\!\left(\!\frac{D_y}{2},\! \sqrt{C_5},\!\frac{D_y}{2}\right)$}
            \\ \cline{5-6}
            &&&&
            $C_9<0$&
            $1-\Phi_2\left(\sqrt{C_5},\sqrt{C_6}\right)-\Phi_3\left(\frac{D_y}{2},\sqrt{C_7},\frac{D_y}{2}\right)$
            \\ \cline{4-6}
            &&&
            \multirow{3}{*}[0cm]{$C_5<C_6$}&
            $C_9\geq 0,C_9<C_7$&
            $1-\Phi_1\left(\sqrt{C_9},\sqrt{C_6}\right)-\Phi_3\left(\frac{D_y}{2},\sqrt{C_7},\frac{D_y}{2}\right)$
            \\ \cline{5-6}
            &&&&
            $C_9\geq 0,C_9\geq C_7$&
            $1-\Phi_1\left(\sqrt{C_7},\sqrt{C_6}\right)-\Phi_3\left(\frac{D_y}{2},\sqrt{C_7},\frac{D_y}{2}\right)$
            \\ \cline{5-6}
            &&&&
            $C_9<0$&
            $1-\Phi_3\left(\frac{D_y}{2},\sqrt{C_7},\frac{D_y}{2}\right)$
            \\ \hline \hline
        \end{tabular}
    \end{table*}
    \begin{table*}
        \caption{Expressions for OP $x_{2a}$ $C_6\geq0,\sqrt{C_6}<\frac{D_y}{2}$}
        \label{OP2av6}
        \centering
        \begin{tabular}{c|c|c|c|c|c}
            \hline \hline
            \multicolumn{5}{c|}{Conditions}
            & Expression
            \\ \hline \hline
            \multirow{29}{*}[0cm]{\rotatebox{90}{$\sqrt{C_7} \geq \frac{D_y}{2}$}} &
            $C_8<C_6$&
            &
            &
            &
            $1$
            \\ \cline{2-6}
            &
            \multirow{14}{*}[0cm]{\rotatebox{90}{$C_8\geq C_6, \sqrt{C_8}<\frac{D_y}{2}$}}&
            $C_4<C_6$&&&
            $1$
            \\ \cline{3-6}
            &&
            \multirow{6}{*}[0cm]{\rotatebox{90}{$\!C_4\!\!\geq\!\! C_6,\! C_4\!<\! C_8$}}&
            \multirow{3}{*}[0cm]{$C_5\geq C_6,C_5<C_8$}&
            $C_9\geq0,C_9<C_4$&
            $1-\Phi_2\left(\sqrt{C_5},\sqrt{C_6}\right)-\Phi_1\left(\sqrt{C_9},\sqrt{C_5}\right)$
            \\ \cline{5-6}
            &&&&
            $C_9\geq0,C_9\geq C_4$&
            $1-\Phi_2\left(\sqrt{C_5},\sqrt{C_6}\right)-\Phi_1\left(\sqrt{C_4},\sqrt{C_5}\right)$
            \\ \cline{5-6}
            &&&&
            $C_9<0$&
            $1-\Phi_2\left(\sqrt{C_5},\sqrt{C_6}\right)$
            \\ \cline{4-6}
            &&&
            \multirow{3}{*}[0cm]{$C_5<C_6$}&
            $C_9\geq 0,C_9<C_4$&
            $1-\Phi_1\left(\sqrt{C_9},\sqrt{C_6}\right)$
            \\ \cline{5-6}
            &&&&
            $C_9\geq 0,C_9\geq C_4$&
            $1-\Phi_1\left(\sqrt{C_4},\sqrt{C_6}\right)$
            \\ \cline{5-6}
            &&&&
            $C_9< 0$&
            $1$
            \\ \cline{3-6}
            &&
            \multirow{7}{*}[0cm]{\rotatebox{90}{$C_4\geq C_8$}}&
            $C_5\geq C_8$&&
            $1-\Phi_2\left(\sqrt{C_8},\sqrt{C_6}\right)$
            \\ \cline{4-6}
            &&&
            \multirow{3}{*}[0cm]{$C_5\geq C_6,C_5<C_8$}&
            $C_9\geq0,C_9<C_8$&
            $1-\Phi_2\left(\sqrt{C_5},\sqrt{C_6}\right)-\Phi_1\left(\sqrt{C_9},\sqrt{C_5}\right)$
            \\ \cline{5-6}
            &&&&
            $C_9\geq0,C_9\geq C_8$&
            $1-\Phi_2\left(\sqrt{C_5},\sqrt{C_6}\right)-\Phi_1\left(\sqrt{C_8},\sqrt{C_5}\right)$
            \\ \cline{5-6}
            &&&&
            $C_9<0$&
            $1-\Phi_2\left(\sqrt{C_5},\sqrt{C_6}\right)$
            \\ \cline{4-6}
            &&&
            \multirow{3}{*}[0cm]{$C_5<C_6$}&
            $C_9\geq0,C_9<C_8$&
            $1-\Phi_1\left(\sqrt{C_9},\sqrt{C_6}\right)$
            \\ \cline{5-6}
            &&&&
            $C_9\geq0,C_9\geq C_8$&
            $1-\Phi_1\left(\sqrt{C_8},\sqrt{C_6}\right)$
            \\ \cline{5-6}
            &&&&
            $C_9<0$&
            $1$
            \\ \cline{2-6}
            &
            \multirow{14}{*}[0cm]{\rotatebox{90}{$\sqrt{C_8}\geq \frac{D_y}{2}$}}&
            $C_4<C_6$&&&
            $1$
            \\ \cline{3-6}
            &&
            \multirow{6}{*}[0cm]{\rotatebox{90}{$ C_4\!\!\geq\!\! C_6,\! \sqrt{C_4}\!\!<\!\!\frac{D_y}{2}$}}&
            \multirow{3}{*}[0cm]{$C_5\geq C_6, \sqrt{C_5}<\frac{D_y}{2}$}&
            $C_9\geq 0,C_9<C_4$&
            $1-\Phi_2\left(\sqrt{C_5},\sqrt{C_6}\right)-\Phi_1\left(\sqrt{C_9},\sqrt{C_5}\right)$
            \\ \cline{5-6}
            &&&&
            $C_9\geq 0,C_9\geq C_4$&
            $1-\Phi_2\left(\sqrt{C_5},\sqrt{C_6}\right)-\Phi_1\left(\sqrt{C_4},\sqrt{C_5}\right)$
            \\ \cline{5-6}
            &&&&
            $C_9<0$&
            $1-\Phi_2\left(\sqrt{C_5},\sqrt{C_6}\right)$
            \\ \cline{4-6}
            &&&
            \multirow{3}{*}[0cm]{$C_5<C_6$}&
            $C_9\geq 0,C_9<C_4$&
            $1-\Phi_1\left(\sqrt{C_9},\sqrt{C_6}\right)$
            \\ \cline{5-6}
            &&&&
            $C_9\geq 0,C_9\geq C_4$&
            $1-\Phi_1\left(\sqrt{C_4},\sqrt{C_6}\right)$
            \\ \cline{5-6}
            &&&&
            $C_9<0$&
            $1$
            \\ \cline{3-6}
            &&
            \multirow{7}{*}[0cm]{\rotatebox{90}{$\sqrt{C_4}\geq\frac{D_y}{2}$}}&
            $\sqrt{C_5}\geq\frac{D_y}{2}$&&
            $1-\Phi_2\left(\frac{D_y}{2},0\right)$
            \\ \cline{4-6}
            &&&
            \multirow{3}{*}[0cm]{$C_5\geq C_6,\sqrt{C_5}<\frac{D_y}{2}$}&
            $C_9\geq0,\sqrt{C_9}<\frac{D_y}{2}$&
            $1-\Phi_2\left(\sqrt{C_5},\sqrt{C_6}\right)-\Phi_1\left(\sqrt{C_9},\sqrt{C_5}\right)$
            \\ \cline{5-6}
            &&&&
            $C_9\geq0,\sqrt{C_9}\geq\frac{D_y}{2}$&
            $1-\Phi_2\left(\sqrt{C_5},\sqrt{C_6}\right)-\Phi_1\left(\frac{D_y}{2},\sqrt{C_5}\right)$
            \\ \cline{5-6}
            &&&&
            $C_9<0$&
            $1-\Phi_2\left(\sqrt{C_5},\sqrt{C_6}\right)$
            \\ \cline{4-6}
            &&&
            \multirow{3}{*}[0cm]{$C_5<C_6$}&
            $C_9\geq0,\sqrt{C_9}<\frac{D_y}{2}$&
            $1-\Phi_1\left(\sqrt{C_9},\sqrt{C_6}\right)$
            \\ \cline{5-6}
            &&&&
            $C_9\geq0,\sqrt{C_9}\geq\frac{D_y}{2}$&
            $1-\Phi_1\left(\frac{D_y}{2},\sqrt{C_6}\right)$
            \\ \cline{5-6}
            &&&&
            $C_9<0$&
            $1$
            \\ \hline \hline
        \end{tabular}
    \end{table*}
\begin{IEEEproof}
    To successfully decode message $x_{2a}$, it is required that the other messages, i.e., $x_{1a}$, $x_{b}$, have also been successfully decoded. Thus, a similar rationale to the previous proof is followed. Specifically, the probability of decoding all transmitted messages successfully is calculated and then the complementary event is considered. Taking this into account, we have 
    \begin{equation} \small
        P_{s,x_{2a}}=\Pr\left(\underbrace{\gamma_b\geq \theta_2}_{E_1},\underbrace{\gamma_{1a}\geq \theta_{11}}_{E_2}, \underbrace{\gamma_{2a}\geq\theta_{12}}_{E_3}\right).
    \end{equation}
Regarding $E_3$, we set $P_3 = \Pr\left(E_3\right)$, which is given by
\begin{equation} \small
    P_3=\Pr\left(y_{U,1}^2\leq \frac{(1-\alpha)\eta\gamma_1}{\theta_{12}}-d^2=C_8\right),
\end{equation}
while the rest have been analyzed in the proof in Section \ref{ProofXb}. Considering all the possible values of these parameters and how they compare with each other and with the dimensions of the room, i.e., $0$ and $\frac{D_y}{2}$, the various cases arise.  
\end{IEEEproof}
\begin{remark}
    Considering \eqref{Pout1A}, it is obvious that the right-hand side of the inequality is positive. Thus, to successfully decode message $x_{1a}$, both terms in the left-hand side must be positive. Taking this into account, it is required that $\frac{\alpha}{1-\alpha}\geq \theta_{11}$. This is a crucial insight into the system design, since being unable to decode the first message results in an outage for the rest of the messages as well. 
\end{remark}
\begin{remark}
    It should be highlighted that in these tables no expressions for $\sqrt{C_6}\geq \frac{D_y}{2}$ appear. This can be explained by considering that $y_{U,1}\in[0,\frac{D_y}{2}]$, which means that $y_{U,1}<\sqrt{C_6}, \forall y_{U,1}$. Thus, $C_2<0$, making the inequality $y_{U,2}^2\leq C_2$ infeasible, resulting in an outage always happening. Taking this into account, $P_{o,x_{b}}=P_{o,x_{2a}}=1$, when  $\sqrt{C_6}\geq \frac{D_y}{2}$.
\end{remark}
\begin{remark}
    It should be noted that $C_4<0$ results in $P_{o,x_{b}}=P_{o,x_{2a}}=1$. This is based on the fact that $y_{U,1}^2 \geq C_4$ results in $\sqrt{C_1}\geq \frac{D_y}{2}$. Considering that $y_{U,1}^2\geq0$, having $C_4<0$ results in $\sqrt{C_1}\geq \frac{D_y}{2}, \forall y_{U,1}$ making the inequality $C_1 \leq y_{U,2}^2$  infeasible $\forall y_{U,2}\in[0,\frac{D_y}{2}]$. Thus, the constant OP is explained.
\end{remark}
    \section{Numerical Results and Simulations}
    In this section, the performance of the considered network is evaluated and the theoretical analysis is validated by Monte Carlo simulations with $10^6$ realizations. We assume that the carrier frequency is $f_c = 28$GHz, the effective refractive index $n_{e} = 1.4$ and the height of the room is $d=3$m. It should be highlighted that the performance of the proposed scheme is derived for the optimal values of the power allocation factor $\alpha$ and target rate factor $\beta$.
    
    In Fig. \ref{OPvsSNRRate}, the OP versus the transmit SNR is illustrated for various rate thresholds. The proposed PAS, termed PAS RSMA, is compared with a PAS using NOMA. The dimensions of the room is set at $D_x = 20$m and $D_y = 20$m. As expected, lowering the system's rate requirements results in a significantly lower OP. The same result can be achieved by increasing the transmit SNR. The effect of these changes can be easily explained by considering the expressions used to calculate the OP. Specifically, the first change, i.e., lowering the rate threshold,   lowers the right-hand side of the inequalities, while the second one, i.e., increasing the SNR, increases the left-hand side of the inequalities. Since the other side of the inequality remains constant, both of these changes reduce the probability that the inequalities are true. In addition, while in lower SNR regions the performance of RSMA is comparable to that of NOMA, in higher SNR regions the proposed system substantially outperforms NOMA. Most importantly, due to the flexibility of RSMA regarding the power allocated to each stream, it avoids the OP floors for increased rate thresholds.
    
\begin{figure}
        \centering
        \begin{tikzpicture}
            \begin{semilogyaxis}[
                width = 0.82\linewidth,
                xlabel = {Transmit SNR (dB)},
                ylabel = {Outage Probability},
                ymin = 0.01,
                ymax = 1,
                xmin = 75,
                xmax = 95,
                grid = major,
                legend entries = {{PAS RSMA},{PAS NOMA},{},{},{},{$R_a=R_b=1$},{},{$R_a=R_b=0.5$}},
                legend cell align = {left},
                legend style = {font = \scriptsize},
                legend style={at={(0,0)},anchor=south west}
                ]
                \addplot[
                black,
                mark = square,
                mark repeat = 2,
                mark size = 3,
                mark phase = 0,
                only marks,
                ]
                table {Data/OPvsSNR/RSMARB12020.dat};   
                \addplot[
                black,
                mark = triangle,
                mark repeat = 2,
                mark size = 3,
                mark phase = 0,
                only marks,
                ]
                table {Data/OPvsSNR/NOMARB12020.dat};
                \addplot[
                red,
                mark = square,
                mark repeat = 2,
                mark size = 3,
                mark phase = 0,
                only marks,
                ]
                table {Data/OPvsSNR/RSMAR0.52020.dat};   
                \addplot[
                red,
                mark = triangle,
                mark repeat = 2,
                mark size = 3,
                mark phase = 0,
                only marks,
                ]
                table {Data/OPvsSNR/NOMAR0.52020.dat};
                \addplot[
                black,
                mark = square,
                mark repeat = 10,
                mark size = 3,
                mark phase = 0,
                no marks,
                line width = 1pt
                ]
                table {Data/OPvsSNR/RSMARB12020.dat};   
                \addplot[
                black,
                mark = triangle,
                mark repeat = 10,
                mark size = 3,
                mark phase = 0,
                no marks,
                line width = 1pt
                ]
                table {Data/OPvsSNR/NOMARB12020.dat};
                \addplot[
                red,
                mark = square,
                mark repeat = 2,
                mark size = 3,
                mark phase = 0,
                no marks,
                line width = 1pt
                ]
                table {Data/OPvsSNR/RSMAR0.52020.dat};   
                \addplot[
                red,
                mark = triangle,
                mark repeat = 2,
                mark size = 3,
                mark phase = 0,
                no marks,
                line width = 1pt
                ]
                table {Data/OPvsSNR/NOMAR0.52020.dat};
            \end{semilogyaxis}
        \end{tikzpicture}
        \vspace{-4mm}
        \caption{OP vs transmit SNR for different rate thresholds.}
        \vspace{-2mm}
        \label{OPvsSNRRate}
    \end{figure}
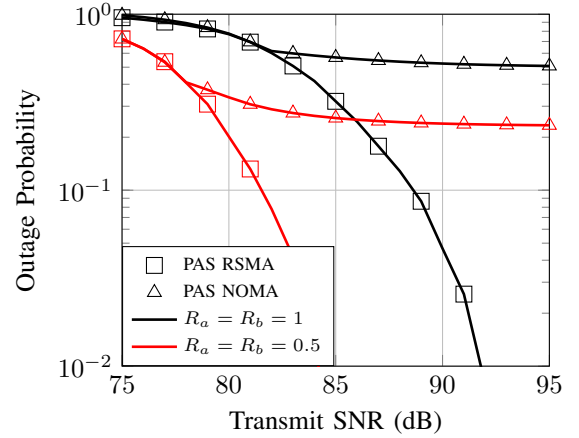 
    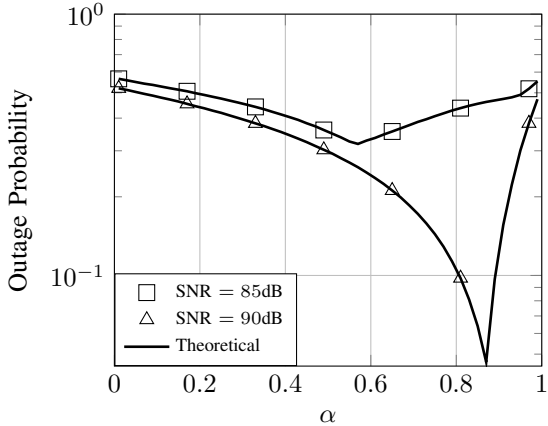
\begin{figure}
        \centering
        \begin{tikzpicture}
            \begin{semilogyaxis}[
            width = 0.82\linewidth,
            xlabel = {$\alpha$},
            ylabel = {Outage Probability},
            ymin = 0.045,
            ymax = 1,
            xmin = 0,
            xmax = 1,
            grid = major,
            legend entries = {{SNR $= 85$dB},{SNR $= 90$dB},{Theoretical}},
            legend cell align = {left},
            legend style = {font = \scriptsize},
            legend style={at={(0,0)},anchor=south west}
            ]
            \addplot[
            black,
            mark = square,
            mark repeat = 8,
            mark size = 3,
            mark phase = 0,
            only marks,
            ]
            table {Data/OPvsA/OPvsASNR85.dat};       
            \addplot[
            black,
            mark = triangle,
            mark repeat = 8,
            mark size = 3,
            mark phase = 0,
            only marks,
            ]
            table {Data/OPvsA/OPvsASNR90.dat};     
            \addplot[
            black,
            mark = square,
            mark repeat = 2,
            mark size = 3,
            mark phase = 0,
            no marks,
            line width = 1pt
            ]
            table {Data/OPvsA/OPvsASNR85.dat};     
            \addplot[
            black,
            mark = square,
            mark repeat = 2,
            mark size = 3,
            mark phase = 0,
            no marks,
            line width = 1pt
            ]
            table {Data/OPvsA/OPvsASNR90.dat};     
            \end{semilogyaxis}
        \end{tikzpicture}
        \vspace{-4mm}
        \caption{OP vs $\alpha$.}
        \vspace{-4mm}
        \label{OPvsA}
    \end{figure}
    Fig. \ref{OPvsA} shows the OP of the system versus the power allocation for message $x_{1a}$, $\alpha$ for two transmit SNR values, $D_y = 20$m and $R_a=R_b=1$. This figure highlights the importance of choosing an optimal $\alpha$, since a suboptimal one could lead to significant system performance degradation. Furthermore, it provides practical insights into the relationship between the optimal value of $\alpha$ and the transmit SNR. Specifically, increasing the SNR results in a higher optimal value for $\alpha$. This can be explained by considering the transition from a noise-limited system to an interference-limited one. In more detail, when the SNR is low, noise is the main cause of outages, thus sufficient transmit power is required for the last message, $x_{2a}$, to be correctly decoded. Conversely, in high SNR scenarios, interference from the other messages leads to outages. Taking this into account, it is optimal to allocate most of the transmit power to the first message, $x_{1a}$, which treats the rest of the messages as interference, to ensure its correct transmission. Due to the increased SNR, the remaining transmit power is sufficient to overcome the impact of noise on decoding message $x_{2a}$. Finally, in agreement with Fig. \ref{OPvsSNRRate}, increasing the SNR leads to a lower OP provided that an optimal $\alpha$ is selected.
    \section{Conclusions}
    In this work, an uplink PAS comprising two PAs and two users implementing RSMA was investigated. Specifically, novel closed-form expressions for the OP, an important metric to evaluate the performance of a system when users use fixed transmission rates, of such a system were derived. These expressions provided useful insights into the practical design of such a system and indicate ways to optimize it. Furthermore, numerical results corroborate the theoretical analysis and explicitly show the impact of each system parameter on its performance. Finally, they demonstrated the superiority of the investigated scheme over PAS NOMA. Taking this into account, this work lays the foundation for further investigation of other performance metrics.
    
    \bibliographystyle{ieeetr}
    \bibliography{Bibliography}
\end{document}